\documentclass[aps,prd,preprint]{revtex4}

\usepackage{graphicx}
\usepackage{psfrag}

\begin{document}
\title{Hadronic production of the $P$-wave excited $B_c$-states ($B_{cJ,L=1}^*$)}
\author{Chao-Hsi Chang$^{1,2}$ \footnote{email:
zhangzx@itp.ac.cn}, Jian-Xiong Wang$^{3}$\footnote{email:
jxwang@mail.ihep.ac.cn} and Xing-Gang Wu$^{3}$\footnote{email:
wuxg@mail.ihep.ac.cn}}
\address{$^1$CCAST
(World Laboratory), P.O.Box 8730, Beijing 100080,
China.\footnote{Not correspondence address.}\\
$^2$Institute of Theoretical Physics, Chinese Academy of Sciences,
P.O.Box 2735, Beijing 100080, China.\\$^3$Institute of High Energy
Physics, P.O.Box 918(4), Beijing 100039, China}

\begin{abstract}
Adopting the complete $\alpha_s^4$ approach of the perturbative
QCD (pQCD) and updated parton distribution functions, we have
estimated the hadronic production of $P$-wave excited $B_c$-states
($B_{cJ,L=1}^*$). In the estimate, special care on the relation of
the production amplitude to the derivative of wave function at
origin of the potential model is payed. For experimental
references, main uncertainties are discussed, and the total cross
sections and the distributions of the production with reasonable
cuts at the energies of Tevatron and LHC are computed and
presented. The results show that $P$-wave production may
contribute to the $B_c$-meson production indirectly by a factor
about $0.5$ of the direct production, and with such a big cross
section, it is worth further to study the possibility to observe
the $P$-wave production itself experimentally.\\


\end{abstract}

\maketitle

\section{Introduction}

$B_c$ meson has been observed experimentally \cite{CDF,D0}, and
the observations are in consistency with the theoretical
prediction. In addition to the experimental progress, as
theoretical studies of the meson $B_c$ are getting deeply and
widely \cite{B-work,mord,qigg,prod0,prod,
prod1,prod2,prod3,prod4,spec,chen,dec,dec1,life,MG,CCWZ},
interesting properties of the meson $B_c$, such as its lifetime,
mass, decay branching ratios etc, are explored more and more
precisely, the $B_c$ mesons can be used for generating tagged
$B_s$ mesons ($B_c$ meson has very big decay branching ratio to
$B_s$ meson) and for studying two heavy flavors $c$ and $\bar{b}$
simultaneously etc, so $B_c$ physics is attracting more and more
attentions. Considering the usages for experimental feasibility
studies, recently we also wrote a generator named BCVEGPY
\cite{wuxglun} for the hadronic production of $B_c$ ($S$-wave),
which is a Fortran program package and is in PYTHIA format
\cite{pythia}.

Since excited states of $B_c$, such as the $P$-wave states
$B_c^*(^1P_1)$ and $B_c^*(^3P_J) (J=1,2,3)$ (for shortening,
hereafter we use the symbol $B_{cJ,L=1}^*$ to denote these four
$P$-wave states) etc, may directly or indirectly (cascade way)
decay to the ground state with almost 100\% possibility via
electromagnetic or hadronic interactions, thus the production of
the excited $B_c$ states can be additional sources of the $B_c$
meson production, i.e. the $B_c$ mesons are produced `indirectly'
but promptly via excited state decay. If the indirect $B_c$
hadronic production and the direct one cannot be discriminated
experimentally, but one would like to understand the production of
the meson $B_c$ well, one certainly needs to know the indirect
production fraction precisely. If one can discriminate the
indirect $B_c$ production and the direct one and, furthermore, the
excited $B_c$ states can be measured via their decay products
exclusively with experimental techniques, then the potential
models for the ($c\bar{b}$) system may have various tests
\cite{spec}, therefore the theoretical estimate about the excited
state production in advance certainly is very useful references
for the experimental measurements. No matter the excited $B_c$
states can be measured exclusively or not, the theoretical
estimate of the production of the excited states is requested. As
known, it has a long story to understand the hadronic production
of $J/\psi$, even now there are still some problems e.g.
experimental data still indicate not to agree with the theoretical
prediction on the polarization of the hadronic produced $J/\psi$
and the discrepancies of the $J/\psi$ production in the B factory
as well. While $B_c$ meson is flavored explicitly, so not only its
hadronic production but also its excited states' hadronic
production are simpler in comparison with the production of the
hidden flavored $(c\bar{c})$ states. To understand the hadronic
$P$-wave production not only is needed for understanding
$(c\bar{b})$-quarkonium ($B_c, B_c^*,B_{cJ,L=1}^*,\cdots$)
production, but also will help to clarify up the situation about
the hadronic production for hidden heavy flavor $(c\bar{c})$
mesons. As a step, to study the production of the $P$-wave $B_c$
excited states and to write the generator accordingly are
certainly interesting.

However, of the excited $B_c$ state hadronic production, only few
authors of Refs.\cite{cheung,berezhnoy,berezhnoy2} have studied
the production of the $P$-wave ones $B_{cJ,L=1}^*$ so far. In
Ref.\cite{cheung}, the hadronic $B_{cJ,L=1}^*$ production is
calculated with the fragmentation approach, which is comparatively
simple and can reach to the leading logarithm (LL) level of
perturbative QCD (pQCD). From the experiences for the $S$-wave
production, the approach can reach to a high accuracy only when
the production is in the region where the transverse momentum
($p_t$) of $B_c$ is very high ($p_t\gg 25$ GeV\cite{prod2}) to
compare with the pQCD $\alpha_s^4$ approach. In addition, the
complete pQCD $\alpha_s^4$ approach has a great advantage from the
experimental point of view, that it retains the information about
the $\bar{c}$ and $b$ quark (jets) associated with the meson in
the production. To retain the information about the $\bar{c}$ and
$b$ quark jets is more relevant experimentally, therefore, the
complete pQCD $\alpha_s^4$ approach to study the $P$-wave
production of the $B_c$ excited states, as the case for $S$-wave
production, is more favored, although it is the lowest order (LO)
PQCD calculation. The authors of Refs.\cite{berezhnoy,berezhnoy2}
adopted the $\alpha_s$ approach and found that the total
cross-section is much greater than that predicted by the
fragmentation approach, and the results from fragmentation
approach can be compared with those from the $\alpha_s^4$ approach
only in the region where the transverse momentum of
$B_{cJ,L=1}^*$, $p_t$, is so high as $p_t\geq 30$ GeV (higher than
$S$-wave production) \cite{berezhnoy}.

The amplitude for the $P$-wave state production involves the
derivative of the wave function, so the derivation for the
$P$-wave production is more complicated than that of the $S$-wave
state production a little. While, of the existence $\alpha_s^4$
approach calculations \cite{berezhnoy,berezhnoy2}, the
`derivative' for the $P$-wave production is done numerically.

Since the calculations are quite complicated, and the authors of
Ref.\cite{berezhnoy} did the necessary derivative in the amplitude
numerically, so to re-calculate and to verify the production
calculations in terms of the $\alpha_s^4$ approach is needed. To
meet the needs, we also pay special attention on establishing the
dependence correctly of the amplitude for the production on the
derivative of the wave function at original which is fully
determined in potential models. In fact, it is also to establish
the correct relation between the matrix element appearing in
nonrelativistic QCD (NRQCD) \cite{nrqcd} and the derivative of the
wave function at original precisely. Therefore, we treat the hard
subprocess amplitude in a strict way, that includes to write the
amplitude under Bethe-Salpeter (BS) equation formulism
\cite{mand}, and to compute the derivative and to simplify the
amplitude of the hard subprocess amplitude analytically as
possible as we can (e.g. to find independent basic fermion strings
and to expand the terms corresponding to Feynman diagrams
accordingly) etc. Finally, the obtained amplitude before it to be
squared is quite `condensed'. To guarantee the rightness of the
calculations, we has done the comparisons between the results of
ours and those in Refs.\cite{berezhnoy,berezhnoy2} with the same
input parameters, and done the checks of the gauge invariance for
the amplitude numerically up to the computer abilities.

The paper is organized: to follow Introduction, in Sec.II we
highlight the dominant mechanism of gluon-gluon fusion, and
present the basic formula for the $B_{cJ,L=1}^*$ hadronic
production in the complete $\alpha_s^4$ approach. In Sec.III, we
describe the input parameters for numerical calculations and
present the results properly. Finally, in Sec.IV  we present
discussions and a short summary.

\section{Formulation and technique}

Based on the factorization theorem of perturbative QCD (pQCD), the
relevant hard subprocess plays a key role for hadronic production.
In hadron collisions at high energies, for the production of the
meson $B_c$ and its excited states, the gluon-gluon fusion
subprocess $gg\rightarrow B_c(B_{cJ,L=1}^*)+b+\bar{c}$ is dominant
\cite{B-work,mord,prod0,prod, prod1,prod2,prod3,prod4}. All of the
discussions and calculations in the paper are based on the
gluon-gluon fusion subprocess.
\begin{figure}
\centering
\includegraphics[width=0.50\textwidth]{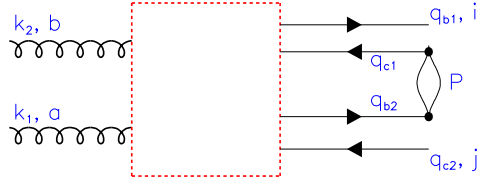}
\vspace{-46mm} \caption{The schematic diagram for the hadronic
production of $B_{cJ,L=1}^*$, where the dashed box stands for the
hard interaction kernel. $k_1$ and $k_2$ are two momenta for the
initial gluons, $q_{b1}$ and $q_{c2}$ are the momenta for the
outgoing b and $\bar{c}$, $P$ is the momentum of $B_{cJ,L=1}^*$.
$i$, $j$, $a$ and $b$ are the color indexes of b, $\bar{c}$ and
the corresponding gluons.} \label{feynshort}
\end{figure}

At the lowest order of pQCD ($\alpha_s^4$ order), there are 36
Feynman diagrams for the gluon-gluon fusion process $gg\rightarrow
B_{cJ,L=1}^* +b+\bar{c}$ totally, which can be schematically
expressed as FIG.\ref{feynshort}, so accordingly there are 36
terms in the production amplitude. As in Ref.\cite{wuxglun}, here
we would like to establish the exact relation of each term of the
amplitude to the wave function (derivative of the wave function at
origin for $P$-wave production) of potential model (also the
relation between the matrix element in non-relativistic QCD
(NRQCD) framework\cite{nrqcd} and the wave function in potential
model framework), thus we start with the `rule' in Ref.\cite{mand}
to write down the explicit expression for the term of the
amplitude corresponding to the $k$th Feynman diagram:
\begin{equation}\label{amp}
M^{[^{(2S+1)}P_J,Jz]}_{k}=C^{ab}_{k,ij}\bar{u}(q_{b1})i\int\frac{d^{4}q}
{(2\pi)^{4}}\left(\Gamma_{1k}\cdot\bar{\chi}^{[^{(2S+1)}P_J,J_z]}(q)\cdot
\Gamma_{2k}\right) v(q_{c2})\;,
\end{equation}
where $\bar{\chi}^{[^{(2S+1)}P_J,J_z]}(q)$ is the Bethe-Salpeter
(BS) wave function for the bound state of ($c\bar{b}$) in
$^{(2S+1)}P_J$, where $C^{ab}_{k,ij}$ is the color factor, $i$,
$j$ are the outgoing $b$ and $\bar{c}$ quarks' color indexes and
$a$, $b$ are the color indexes of the initial two gluons
respectively. $\Gamma_{1k}$ stands for the structure of the $k$-th
Feynman diagram that is between $\bar{u}(q_{b1})$ and the BS wave
function $\bar{\chi}^{[^{(2S+1)}P_J,J_z]}(q)$, which includes the
relative string of Dirac $\gamma$ matrices and the corresponding
scalar part of the propagators, $\Gamma_{2k}$ stands for the
similar structure between the BS wave function
$\bar{\chi}^{[^{(2S+1)}P_J,J_z]}(q)$ and $v(q_{c2})$. $q_{b1}$ and
$q_{c2}$ are the momenta for the outgoing $b$ quark and $\bar{c}$
antiquark. In the non-relativistic approximation, the BS wave
function $\bar{\chi}^{[^{2S+1}P_JJ_z]}(q)$ can be written:
\begin{eqnarray}\label{bound}
&\displaystyle \bar{\chi}^{[^{(2S+1)}P_J,J_z]}(q) \simeq
\sum_{S_z,\lambda,\lambda'}\frac{-\sqrt{M}}{4m_{b}m_{c}}\cdot
\Psi(q)\cdot(\epsilon^\lambda
\cdot q)\cdot\nonumber \\
&(\hat{q}_{b2}-m_{b})\cdot (\delta_{S,0}\delta_{S_z,0}\gamma_5
+\delta_{S,1}\delta_{S_z,\lambda'}\hat\epsilon^{\lambda'})\cdot
(\hat{q}_{c1}+m_{c})\cdot \langle 1\lambda;SS_z |JJ_z \rangle,
\end{eqnarray}
where $P\equiv P_{B_{cJ,L=1}^*}$, $M\equiv m_{B_{cJ,L=1}^*}\simeq
m_c+m_b$ are the momentum, the mass of $B_{cJ,L=1}^*$ meson
respectively (i.e. for shortening the notation, we adopt $P$ and
$M$ to denote the momentum, the mass of $B_{cJ,L=1}^*$ meson
respectively) and $q$ is the relative momentum between the two
constituent quarks. Since $q_{b2}$ and $q_{c1}$ are the momenta of
the $\bar{b}$ and $c$ quarks inside the $B_{cJ,L=1}^*$, so they
relate to the total and relative momenta of the bound state:
\begin{equation}
q_{b2}=\alpha_1 P-q\,,\;\;\; q_{c1}=\alpha_2
P+q\,,\;\;\;\;\alpha_1=\frac{m_{b}}{m_{c}+m_{b}}\,,\;\;\;
\alpha_2=\frac{m_{c}}{m_{c}+m_{b}}\,. \label{eq:momentum}
\end{equation}
For convenience in later usages, let us introduce
$$q^{\mu}_{\perp}\equiv
q^{\mu}-q^\mu_{\parallel}\,,\;\;\;\;\; q^{\mu}_{\parallel}\equiv
\frac{(P\cdot q)}{M^2}P^{\mu}$$ and $q_{\parallel}\equiv
|q^{\mu}_{\parallel}|$, hence accordingly we have
$d^4q=dq_{\parallel}d^{3}q_{\perp}$. The polarization vectors
$$ (\epsilon^\lambda\cdot P)=(\epsilon^{\lambda'}\cdot P)=0
\,,\;\;\;\;\lambda,\;\lambda'=(1,0,-1)\,;$$
$\Psi(q)$ stands for
the `$P$-wave scalar wave function', and $\langle 1\lambda;
SS_z|JJ_z \rangle$ is the Clebsch-Gordon coefficient for L-S
coupling. In the paper, we will use $\hat{a}$ to denote the
contraction between the Dirac $\gamma$ matrix and a momentum or
polarization vector $a$, i.e. we use $\hat{a}$ instead of
$\slash\!\!\!a$. The spin structure of the BS wave function
$\bar{\chi}^{[^{(2S+1)}P_J,J_z]}(q) $ defined in Eq.(\ref{bound})
is of the lowest order (up to ${\cal O}(q)$) for the $P$-wave
state production.

The $P$-wave scalar function $\Psi(q)$ in Eq.(\ref{bound}), as a
factor of BS wave function, relates to the derivative of the wave
function at origin $\psi'_{0}(0)$ in coordinate representation
under the so-called instantaneous approximation \cite{salp} by the
integration:
\begin{eqnarray}
\label{pwavezero}
i\int\frac{dq_{\parallel}d^{3}q_{\perp}}{(2\pi)^{4}}q^{\alpha}
\Psi(q)(\epsilon^\lambda \cdot
q)=i\int\frac{d^{3}q_{\perp}}{(2\pi)^{3}}\tilde{q}_{\perp}^{\alpha}
\phi(-\frac{q^2_{\perp}}{M^2})(\epsilon^\lambda \cdot
q_{\perp})=i\epsilon^{\lambda\;\alpha} \psi^{\prime}(0)\,.
\end{eqnarray}
Here ${\tilde{q}}_{\perp}$ is the unit vector
$\frac{q_{\perp}}{\sqrt{-q^2_{\perp}}}$, and
$\phi(-\frac{q^2_{\perp}}{M^2})$ stands for the $P$-wave `scalar
wave function' in the sense of potential model (the two quarks, as
components of the bound state, must be in a relative space-like
distance, hence $\phi$ should be a function of the variable
$\frac{q^2_{\perp}}{M^2}$). Substituting Eq.(\ref{eq:momentum})
into Eq.(\ref{bound}), we obtain
\begin{eqnarray}\label{eq:ab}
i\int\frac{dq_{\parallel}}{(2\pi)}\bar{\chi}^{[^{(2S+1)}P_J,J_z]}(q)&\simeq
&\sum_{S_z,\lambda,\lambda'}\phi(-\frac{q^2_{\perp}}{M^2})(\epsilon^\lambda
\cdot q_{\perp})\langle 1\lambda ;SS_z | JJ_z \rangle \nonumber \\
&\cdot&\Big\{\frac{1}{2\sqrt{M}}(-\hat{P}+M)\cdot
(\delta_{S,0}\delta_{S_z,0}\gamma_5
+\delta_{S,1}\delta_{S_z,\lambda'}\hat\epsilon^{\lambda'})\nonumber\\
&-&\Big(\frac{\sqrt{M}}{4m_bm_c}\Big) \cdot
\Big[\alpha_2\hat{q_{\perp}}
(-\hat{P}+M)(\delta_{S,0}\delta_{S_z,0}\gamma_5
+\delta_{S,1}\delta_{S_z,\lambda'}\hat\epsilon^{\lambda'})\nonumber\\
&+&\alpha_1(-\hat{P}+M)\cdot(\delta_{S,0}\delta_{S_z,0}\gamma_5
+\delta_{S,1}\delta_{S_z,\lambda'}\hat\epsilon^{\lambda'})\hat{q_{\perp}}\Big]
+{\cal O}(q_{\perp}^2)\Big\}\nonumber\\
&=&\sum_{S_z,\lambda,\lambda'}\phi(-\frac{q^2_{P\perp}}{M^2})
(\epsilon^\lambda \cdot q_{\perp}) \langle 1\lambda ;SS_z | JJ_z
\rangle (A^{\lambda'}_{SS_z}
+B^{\lambda'\;\mu}_{SS_z}q_{{\perp}\mu}\nonumber \\
&+&{\cal O}(q_{\perp}^2)),
\end{eqnarray}
where
\begin{eqnarray}
&\displaystyle A^{\lambda'}_{SS_z}\equiv
\frac{1}{2\sqrt{M}}(-\hat{P}+M)
\cdot(\delta_{S,0}\delta_{\lambda',0}\gamma_5
+\delta_{S,1}\delta_{S_z,\lambda'}\hat\epsilon^{\lambda'})\,,\nonumber\\
&\displaystyle B^{\lambda'\;\mu}_{SS_z}\equiv
-\left(\frac{\sqrt{M}}{4m_bm_c}\right) \cdot
\Big[\alpha_2\gamma^{\mu}
(-\hat{P}+M)\cdot(\delta_{S,0}\delta_{\lambda',0}\gamma_5
+\delta_{S,1}\delta_{S_z,\lambda'}\hat\epsilon^{\lambda'})
\nonumber \\
&+\alpha_1(-\hat{P}+M)\cdot(\delta_{S,0}\delta_{\lambda',0}\gamma_5
+\delta_{S,1}\delta_{S_z,\lambda'}\hat\epsilon^{\lambda'})
\gamma^{\mu}\Big]\,.\nonumber
\end{eqnarray}
Note that at the lowest order relativistic approximation the terms
proportional to $B^{\lambda'\;\mu}_{SS_z}$ do not contribute to
the production of the $S$-wave $B_c (B_c^*)$ meson at all,
however, to the production of the $P$-wave states $B_{cJ,L=1}^*$,
they do contribute, so we have to keep the terms with care. To
reach to the lowest order of the relativistic approximation,
according to Eq.(\ref{amp}) the next step is to do the expansion
of the $\Gamma_{1k}$ and $\Gamma_{2k}$ about $q_\mu$ up to ${\cal
O}(q^2)$ for the $P$-wave production, i.e.
\begin{equation}
\Gamma_{1k}=\Gamma^0_{1k}+\Gamma^{\mu}_{1k}\cdot q_{\mu}+{\cal
O}(q^2)\,,\;\;\;\;\;
\Gamma_{2k}=\Gamma^0_{2k}+\Gamma^{\mu}_{2k}\cdot q_{\mu}+{\cal
O}(q^2)\,,
\end{equation}
where $\Gamma^0_{1k}$, $\Gamma^{\mu}_{1k}$, $\Gamma^0_{2k}$ and
$\Gamma^{\mu}_{2k}$ does not depend on $q^\mu$ (or $q_\mu$) at
all. Substituting the above equations into Eq.(\ref{amp}) and
carrying out the integration over
$d^4q=dq_{\parallel}d^{3}q_{\perp}$ with the help of
Eq.(\ref{pwavezero}), we obtain
\begin{eqnarray}\label{eq:m}
&\displaystyle M_{k}^{S,JJ_z}=
C^{ab}_{k,ij}\psi'(0)\sum_{S_z,\lambda,\lambda'}\langle
1\lambda;SS_z |JJ_z \rangle \epsilon^\lambda_{\mu}\cdot\nonumber\\
&\bar{u}_{s}(q_{b1})\left( \Gamma^{\mu}_{1k}\cdot
A^{\lambda'}_{SS_z}\cdot \Gamma^0_{2k}+ \Gamma^{0}_{1k}\cdot
A^{\lambda'}_{SS_z}\cdot \Gamma^{\mu}_{2k} + \Gamma^{0}_{1k}\cdot
B^{\lambda'\mu}_{SS_z}\cdot \Gamma^0_{2k}\right)
v_{s^{\prime}}(q_{c2}).
\end{eqnarray}
For a specific $P$-wave state, summing over the explicit $\lambda$
and $\lambda'$ for the Clebsch-Gordon coefficients,
Eq.(\ref{eq:m}) can be further simplified as
\begin{eqnarray}\label{eq:simmatrix}
&M_{k}^{S,JJ_z}=C^{^{2S+1}P_J}\psi'(0)
C^{ab}_{k,ij}\nonumber \\
&\cdot\bar{u}(q_{b1}) \left( \Gamma^{\mu}_{1k}\cdot
E_{\mu}^{[^{2S+1}P_J]J_z}\cdot \Gamma^0_{2k}+ \Gamma^{0}_{1k}\cdot
E_{\mu}^{[^{2S+1}P_J]J_z}\cdot \Gamma^{\mu}_{2k} +
\Gamma^{0}_{1k}\cdot F^{[^{2S+1}P_J]J_z}\cdot \Gamma^0_{2k}\right)
v(q_{c2}),
\end{eqnarray}
where the overall factor $C^{^{(2S+1)}P_J}$ and the functions
$E_{\mu}^{[^{2S+1}P_J]J_z}$ and $F^{[^{2S+1}P_J]J_z}$ are computed
separately for different $P$-wave states $B_{cJ,L=1}^*$.

For $^1P_1$ state, $J=1\,,\;J_z=\lambda$ and $S=0$, we obtain
\begin{equation}
E^{[^1P_1]\lambda}_{\mu}=(-\hat{P}+M)\gamma_5\epsilon^\lambda_{\mu}\,,\;\;\;\;
F^{[^1P_1]\lambda}=\frac{-\hat{P}\hat{\epsilon}^\lambda\gamma_5+M(\alpha_1-
\alpha_2)\hat{\epsilon}^\lambda\gamma_5}{2M\alpha_1\alpha_2}\,,\;\;\;\;
C^{^1P_1}=\frac{1}{2\sqrt{M}\sqrt{N_c}}.
\end{equation}
Here and hereafter an extra factor $1/\sqrt{N_c}$ will be
implicitly included due to the fact that the meson is in
color-singlet state.

The coefficients for the $^3P_J (J=0,1,2)$ states with $S=1$ can
be obtained with the help of the relations \cite{chen1,guberina}:
\begin{eqnarray}
\sum_{\lambda,\lambda'}\langle
1\lambda;1\lambda'|00\rangle\epsilon^\lambda_{\mu}
\epsilon^{\lambda'}_{\nu}&=&\sqrt{\frac{1}{3}}\left(\frac{P_{\mu}
P_{\nu}}{P^{2}}-g_{\mu\nu}\right)\,,\;\;\; (J=0)\,,\\
\sum_{\lambda,\lambda'}\langle
1\lambda;1\lambda'|1J_{z}\rangle\epsilon^\lambda_{\mu}
\epsilon^{\lambda'}_{\nu}&=&i\sqrt{\frac{1}{2}} \epsilon_{\mu
\nu}^{\;\;\;\;\;\rho \varrho}\frac{P_{\rho}}{M}
\epsilon_{\varrho}^{J_{z}}\,,\;\;\; (J=1, J_z=-1,0,1)\,,\\
\label{3p2} \sum_{\lambda,\lambda'}\langle
1\lambda;1\lambda'|2J_{z}\rangle\epsilon^\lambda_{\mu}
\epsilon^{\lambda'}_{\nu}&=&\epsilon_{\mu\nu}^{J_{z}}\,,\;(J=2,
J_z=-2,-1,0,1,2)\,,
\end{eqnarray}
and the polarization vector $\epsilon^\lambda_{\mu}$ and tensor
$\epsilon_{\mu\nu}(J_z)$ obey the projection relations
\begin{eqnarray}
\sum_{\lambda}\epsilon^\lambda_{\mu}\epsilon^\lambda_{\nu}&=&\left(\frac{P_{\mu}
P_{\nu}}{P^{2}}-g_{\mu\nu}\right)\equiv \cal{P}_{\mu\nu}\,,\\
\sum_{J_{z}}\epsilon_{\mu\nu}^{J_z}\epsilon_{\rho
\varrho}^{J_z}&=&{\frac{1}{2}
[\cal{P}_{\mu\rho}\cal{P}_{\nu\varrho}+
\cal{P}_{\nu\rho}\cal{P}_{\mu\varrho}]} - {\frac{1}{3}
\cal{P}_{\mu\nu}\cal{P}_{\rho\varrho}}\,,
\end{eqnarray}
for $^3P_0$ state ($\lambda=0$):
\begin{equation}
E^{[^3P_0]\lambda}_{\mu}=(-\hat{P}+M)(M\gamma_{\mu}+P_{\mu})\,,\;\;
F^{[^3P_0]\lambda}=-\frac{3[(\alpha_2-
\alpha_1)\hat{P}+M]}{2\alpha_1\alpha_2}\,,\;\;
C^{^3P_0}=\frac{1}{2\sqrt{3N_c}M^{3/2}}\,;
\end{equation}
for $^3P_1$ state ($\lambda=-1,0,1$):
\begin{equation}
E^{[^3P_1]\lambda}_{\mu}=i(\hat{P}+M)\epsilon_{\mu}^{\;\;\;\rho\varrho\nu}
\gamma_{\rho}P_{\varrho}\epsilon^\lambda_{\nu}\,,\;\;
F^{[^3P_1]\lambda}=-\frac{\hat{P}\hat{\epsilon}^\lambda\gamma_5-
M\hat{\epsilon}^\lambda\gamma_5}{\alpha_1\alpha_2}\,,\;\;
C^{^3P_1}=\frac{1}{2\sqrt{2N_c}M^{3/2}}\,.
\end{equation}
Finally for $^3P_2$ state ($J_z=-2,-1,\cdots,2$):
\begin{equation}
E^{[^3P_2]J_z}_{\mu}=(-\hat{P}+M)\gamma_5\gamma_{\nu}\epsilon_{\mu\nu}^{J_z}\,,
\;\;\; F^{[^3P_2]J_z}=0\,,\;\;\;
C^{^3P_2}=\frac{1}{2\sqrt{N_cM}}\,.
\end{equation}

If one calculates cross sections of the production by summing all
the 36 terms (corresponding to the Feynman diagrams), and then
having the result squared directly, the squared amplitude will be
too long to deal with, even in terms of computer, and is really
time-consuming to reach to the final results for the cross
sections. Instead of summing the terms directly, we adopt the FDC
package\cite{fdc} to generate the Fortran program for each term
and develop a technique in FDC further. Namely we firstly
establish a complete set of ` basic spinor lines', such as that
they are constructed by the multiplication of Dirac $\gamma$
matrixes in a certain way as $\hat p_1\hat p_2 \hat p_3 \cdots$,
then we consider them as bases to expand every terms of the
amplitude and sum up all the terms according to the expansion (the
coefficients of the `bases' are summed respectively). In this way,
the amplitude and its square become quite condense, thus the
efficiency for computing the amplitude squared is raised greatly.

In order to simplify the amplitude manipulation under Fortran
program, it is better to adopting the linear polarizations instead
of the circular ones because the linear ones are real numbers.
Thus we do so, and for convenience we put the explicit form of the
linear polarization vector $\epsilon_{\mu}$ and that of tensor
$\epsilon_{\mu\nu}$ in Appendix I, which are used in our
calculations when summing up the final polarizations of
$B_{c,L+1}^*$.

The color factors for the terms corresponding to the Feynman
diagrams should be also treated correctly. For the gluon-gluon
fusion subprocess, we choose the three independent color
factors\footnote{In Refs.\cite{prod1,prod2}, the number $N$ for
color group is considered as a variable, so it is to conclude that
there are 5 independent color factors there. While here $N$ is
fixed to be equal to 3, thus there are three independent color
factor only.} as follows:
\begin{eqnarray}\label{color1}
T^{1,ab}_{ij}&=&\frac{3\sqrt{26}}{26\sqrt{15}}(2(T^aT^b)_{ij}+
2(T^bT^a)_{ij}-3\delta_{ij}Tr[T^aT^b])\,,\\\label{color2}
T^{2,ab}_{ij}&=&\frac{\sqrt{2}}{2\sqrt{429}}(-11(T^aT^b)_{ij}+
2(T^bT^a)_{ij}-3\delta_{ij}Tr[T^aT^b])\,,\\\label{color3}
T^{3,ab}_{ij}&=&\frac{\sqrt{2}}{2\sqrt{33}}(-3(T^bT^a)_{ij}-
\delta_{ij}Tr[T^aT^b])\,,
\end{eqnarray}
with the normalization:
$\sum\limits_{a,b,i,j}(T^{m,ab\dag}_{ij}T^{n,ab}_{ij})=\delta^{mn}$
($m=1,2,3$). All the color factors $C^{ab}_{k,ij}$ in
Eq.(\ref{amp}) can be expanded as the linear combination of these
three independent color factors. Therefore the amplitude can be
grouped into three terms by `color',
\begin{equation}
M^{S,JJ_z}=\sum_{k=1,\cdots,36}M_k^{S,JJ_z}=\sum_{m=1,2,3}(T^{m,ab}_{ij}
M_{m}^{S,JJ_z})\,.
\end{equation}
For various un-polarized cross sections of the subprocess (in
given $J$ and $S$), we have the square of the amplitude:
\begin{equation}\label{square}
|M^{S,J}|^2=\sum_{J_z}|M^{S,JJ_z}|^2\;.
\end{equation}

According to pQCD factorization theorem, the hadronic production
cross section is formulated as
\begin{equation}
d\sigma=\sum_{ij}\int dx_{1}\int
dx_{2}F^{i}_{H_{1}}(x_{1},\mu^2_{F})\times
F^{j}_{H_{2}}(x_{2},\mu^2_{F})d\hat{\sigma}_{ij\rightarrow
B_{cJ,L=1}^*X}(x_{1},x_{2},\mu^2_{F},\mu^2,Q^2)\,, \label{pqcdf}
\end{equation}
where $F^{i}_{H_1}(x,\mu^2_{F})\,,\;F^{j}_{H_2}(x,\mu^2_{F})$ are
the parton distribution functions (PDFs) of the $i$ and $j$
partons in the hadrons $H_1,\;H_2$ respectively. $\mu^2$ is the
`energy scale squared' where renormalization for the subprocess is
made; $Q^2$ is the `characteristic energy scale of the subprocess
squared' i.e. when setting $\mu^2=Q^2$, then LO pQCD calculations
for the subprocess can obtain the best results, thus without
emphasis $\mu^2=Q^2$ is always set; and $\mu^2_F$ is the `energy
scale squared' where the factorization about the PDFs and the hard
subprocess is made. Usually for LO to obtain the best results, the
factorization and `renormalization' are carried out at the same
energy scale i.e. $\mu_F^2=Q^2$, thus later on we take
$\mu^2=\mu_F^2=Q^2$ and define $d\hat{\sigma}_{ij\rightarrow
B_{cJ,L=1}^*X}(x_{1},x_{2},Q^2)\equiv d\hat{\sigma}_{ij\rightarrow
B_{cJ,L=1}^*X}(x_{1},x_{2},\mu^2_{F},\mu^2,Q^2)$ except one case
when estimating the uncertainty from LO and the ambiguity of the
choices about $\mu_F^2$ and $\mu^2=Q^2$.

Having all the above preparations, we base on Eq.(\ref{pqcdf})
technically to write a program for numerically computing the
hadronic production cross sections at various energies
\cite{bcvegpy2.0}.

\section{Numerical results}

\begin{figure}
\centering
\includegraphics[width=0.46\textwidth]{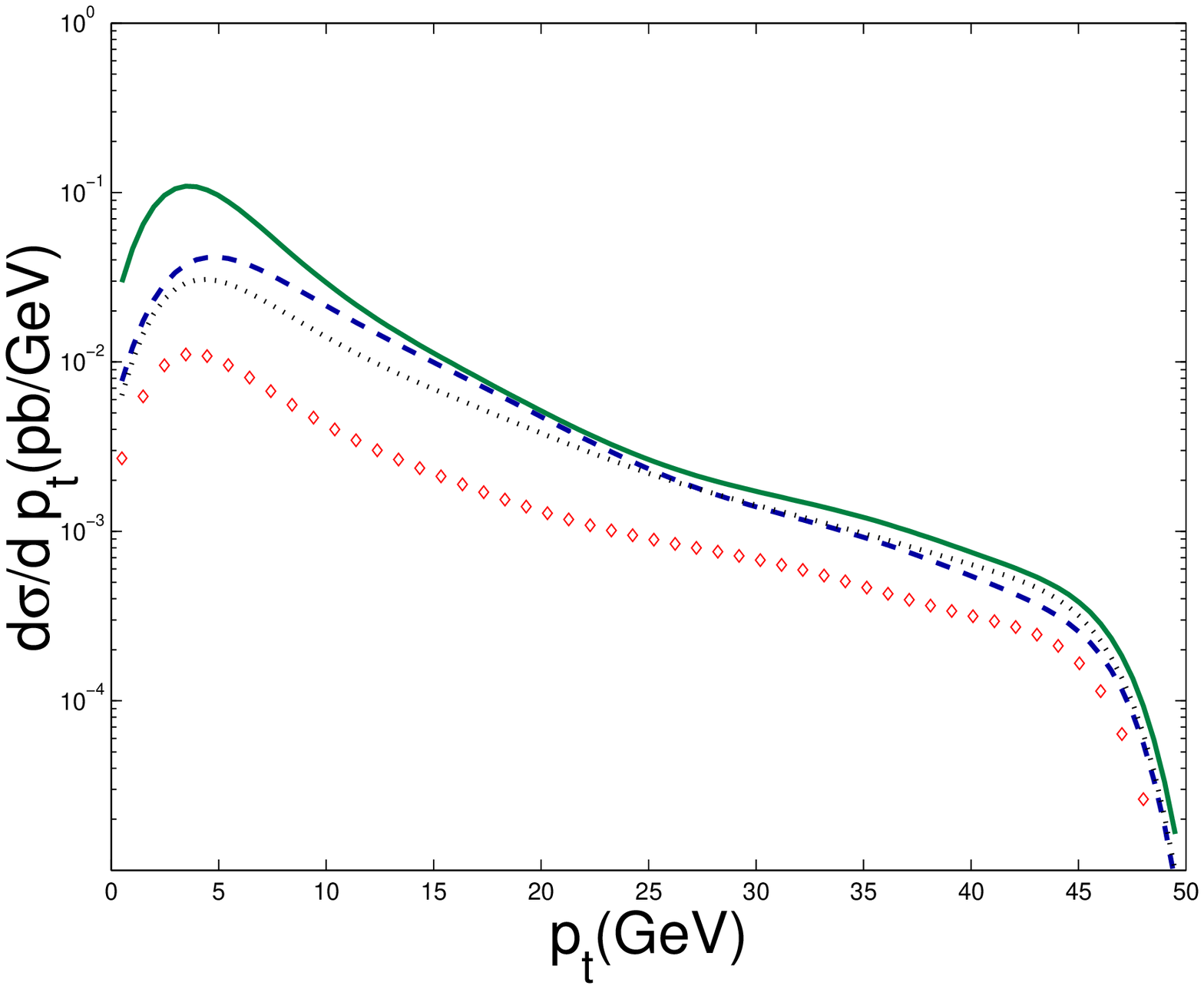}%
\includegraphics[width=0.48\textwidth]{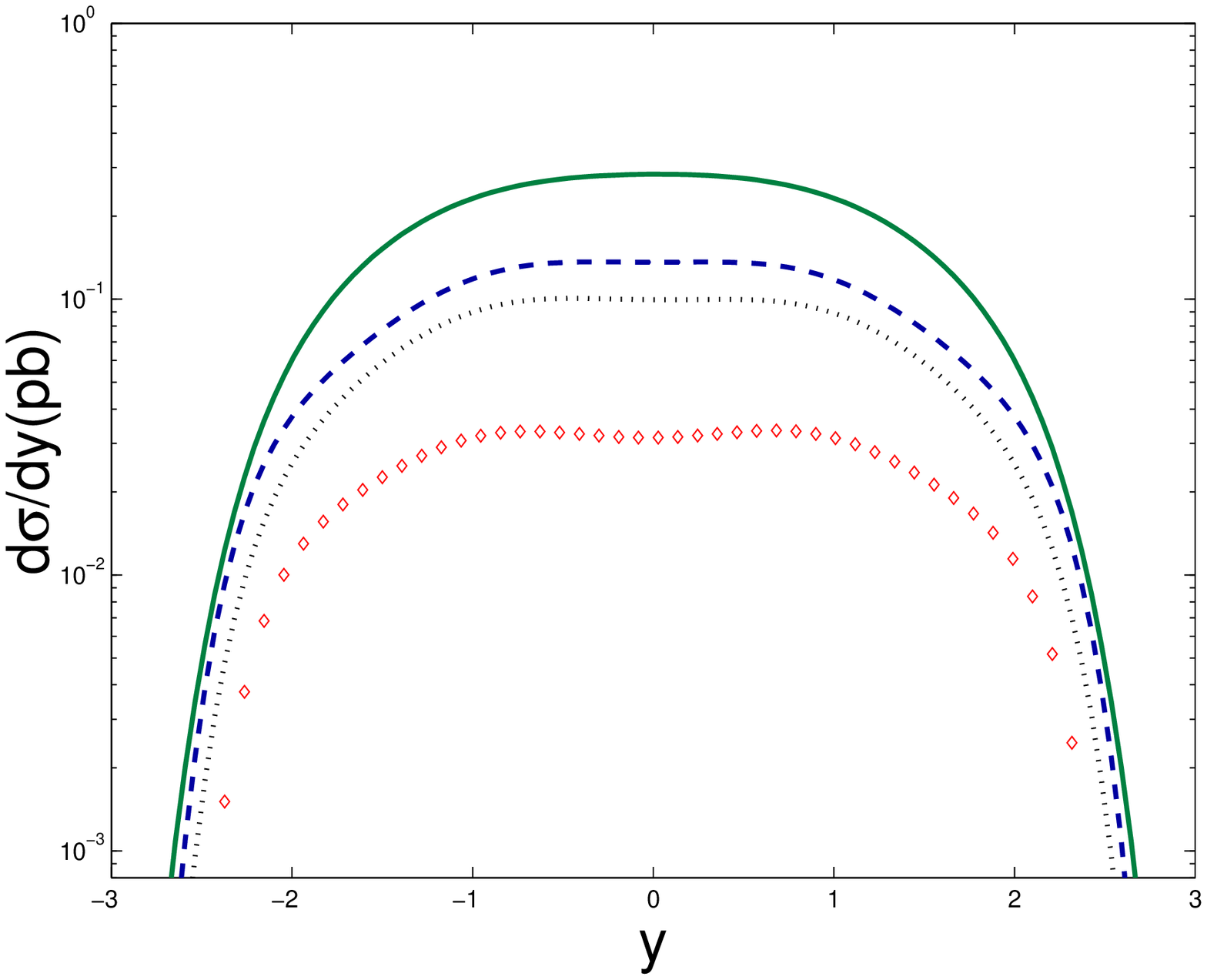}
\caption{\small $B^*_{c,L=1}-p_t$ and $B^*_{c,L=1}-y$ differential
distributions for the subprocess. The dashed line (up-middle), the
diamond line (bottom), the dotted line (down-middle) and the solid
line (top) are for $^1P_1$, $^3P_0$, $^3P_1$ and $^3P_2$,
respectively. The parameters are taken to be the same ones that
have been used in Ref.\cite{berezhnoy}.} \label{subwavep}
\vspace{-4mm}
\end{figure}

Before doing numerical calculations, we have done the necessary
checks. First of all, we numerically check the gauge invariance.
We find the gauge invariance is guaranteed at the used computer
ability level for our formula and computer program to compute the
amplitude. Then we further compute the cross sections numerically
for the subprocess $g+g\rightarrow B_{cJ,L=1}^*+b+\bar{c}$ with
the input parameters as taken in Ref.\cite{berezhnoy}, and the
same numerical results as those in Ref.\cite{berezhnoy} are
obtained (see TABLE \ref{tabsub}, Fig.\ref{subwavep} also
Ref.\cite{berezhnoy}). Thus, we confirm the results of
Ref.\cite{berezhnoy} for the subprocess, and are quite sure the
correctness of our formula and program.

We should note here that the approximation Eq.(\ref{bound}), which
is also taken in Ref.\cite{berezhnoy} and is similar in the cases
for ($c\bar{c}$) and ($b\bar{b}$), the heavy flavor hidden
systems, indicates that the equations $P=q_{c1}+q_{b2}$,
$q_{c1}^2=m_c^2,\;q_{b2}^2=m_b^2$ and $P^2=m_{B_{cJ,L=1}^*}^2$
must be satisfied simultaneously, furthermore, only when the
equations are satisfied, the gauge invariance of the amplitude is
guaranteed. Therefore, under the approximation, there are two
possibilities: one is to take the same quark mass values in the
$P$-wave and $S$-wave state production, but one have to ignore the
mass difference between the $P$-wave and $S$-wave states i.e.
$m_{B_c}=m_{B_{cJ,L=1}^*}$; the other one is to consider the mass
differences between the $P$-wave and $S$-wave states i.e.
$m_{B_c}\neq m_{B_{cJ,L=1}^*}$ in the $P$-wave production, but one
have to take different mass values of $c$ and $b$ quarks from
those in $S$-wave production accordingly. The authors of
Ref.\cite{berezhnoy} took the later i.e. they took $m_b=5.0$ GeV,
$m_c=1.7$ GeV for the $P$-wave state production in order to have a
mass of $B_{cJ,L=1}^*$ about $6.7$ GeV, although they took
$m_{B_c}=6.3$ GeV, $m_c=1.5$ GeV and $m_b=4.8$ GeV for $S$-wave
production in the meantime. As known that the cross section of the
jet production $g+g\rightarrow c+\bar{b}+b+\bar{c}$ is quite
sensitive to the chosen values of the masses $m_b,\;m_c$, thus, we
think that the jet production should not be affected by the
differences for the $P$-wave and $S$-wave production, and in fact
that is also the requirement for the NRQCD formulism, of the two
possibilities which cannot be avoided under the approximation
Eq.(\ref{bound}), the former, i.e. to take
$m_{B_{cJ,L=1}^*}=m_{B_c}$ and to keep the values of $b$ and $c$
quark masses to be the same as those for the $S$-wave production,
should be more relevant comparatively. Hence, in the paper we
carry out all numerical calculations always to take the `choice':
the mass values of $c,b$-quarks as $m_c=1.5$ GeV $m_b=4.9$ GeV,
i.e., the same as those taken in $S$-wave production, so
$m_{B_{cJ,L=1}^*}=m_{B_c}=6.4$ GeV, except the case when showing
the uncertainties from the two possible choices.

As pointed out in Ref.\cite{changwu}, there are quit a lot of
uncertainty sources in the hadronic production for the $S$-wave
$B_c (B_c^*)$ meson, such as the variations about
$\alpha_s$-running, the variations in choices of the factorization
energy scale, the adopted PDFs in different versions and various
input parameter values relating the bound state {\it etc.}. For
instance for definiteness in check of our formula and computer
program by comparison with the results of \cite{berezhnoy}, we
take pQCD coupling constant $\alpha_s=0.2$, the derivative of the
wave function at origin $|R'(0)|^2=0.201 GeV^5$, and the mass
values of $c$-quark, $b$-quarks and $B_{cJ,L=1}^*$ are fixed as
$m_b=5.0GeV$, $m_c=1.7GeV$, $M=6.7GeV$ {\it etc} i.e. all are
fixed as those in \cite{berezhnoy} for the subprocess $gg\to
B_{cJ,L=1}^*+b+\bar{c}$.

\begin{table}
\begin{center}
\caption{The total cross section for the hard subprocess
$g+g\rightarrow B_{cJ,L=1}^*+b+\bar{c}$ (gluon fusion into
$P$-wave excited states $B_{cJ,L=1}^*$) at different C.M.
energies. The input parameters are taken as those used in
Ref.\cite{berezhnoy}: $m_b=5.0GeV$, $m_c=1.7GeV$, $M=6.7GeV$ and
the running $\alpha_s$ is fixed to $0.2$ {\it etc}.}\vspace{4mm}
\begin{tabular}{|c||c|c|c|c|c|c|}
\hline\hline ~C.M. energy (GeV)~ & ~20GeV~ & ~40GeV~ &
 ~60GeV~ & ~80GeV~ & ~100GeV~ & ~200GeV~\\
\hline $\sigma(^1P_1)(pb)$ & 0.184 & 0.743 & 0.657 & 0.538 & 0.439 & 0.195\\
\hline $\sigma(^3P_0)(pb)$ & 0.367 & 0.207 & 0.175 & 0.141 & 0.114 & 0.0496\\
\hline $\sigma(^3P_1)(pb)$ & 0.346  & 0.598 & 0.503 & 0.402 & 0.324 & 0.139\\
\hline $\sigma(^3P_2)(pb)$ & 0.721  & 1.49  & 1.31  & 1.06  & 0.862 & 0.374\\
\hline\hline
\end{tabular}
\label{tabsub}
\end{center}
\end{table}

To compute the production, we need to know the PDFs. Several
groups, such as CTEQ \cite{6lcteq}, GRV \cite{98lgrv} and MRS
\cite{2001lmrst} etc., devote themselves to offer accurate PDFs to
the world. By taking different PDFs, one may find that the results
caused by different PDFs can be about $10\%$. Later on for
definiteness, we will take CTEQ6L\cite{6lcteq} and the leading
order $\alpha_s$ running with $\Lambda^{(n_f=4)}_{QCD}=0.326$ GeV.

As pointed out in Ref.\cite{changwu}, at LO pQCD of all the
uncertainties, that caused by the choice of the energy scale $Q^2$
for the renormalization of the subprocess (or say that for the
factorization) is the largest in the hadronic production estimate,
and for $S$-wave production it may cause such uncertainty as great
around a factor $30\%$. In fact, for $P$-wave production it is
also the case. To see the uncertainty caused by $Q^2$ choice for
the $P$-wave production, we try to calculate the $P$-wave
production under the condition that the other uncertainty factors
are fixed. As in Ref.\cite{changwu}, we take four types of $Q^2$
choices and put the results for total cross section in
Tab.\ref{tabq2} respectively. The four types are:

Type $A$: $Q^2=\hat{s}\,,$ the squared C.M. energy of the
subprocess;

Type $B$: $Q^2=M_t^2\equiv p_{t}^2+m_{B_{cJ,L=1}^*}^2\,,$ the
squared transverse mass of the $B_{cJ,L=1}^*$ meson;

Type $C$, $Q^2=m_{bt}^2\equiv p_{tb}^2+m_b^2\,,$ the squared
transverse mass of the $b$ quark.

Type $D$: $Q^2=4m_b^2$ .

\begin{table}
\begin{center}
\caption{The total cross section (in unit nb) for the hadronic
production of $P$-wave $B_c$ meson ($B_{cJ,L=1}^*$) at LHC and
TEVATRON with different types of factorization energy scales $(A,
B, C, D)$. Here $m_b=4.9$ GeV, $m_c=1.5$ GeV and
$m_{B_{cJ,L=1}^*}=m_c+m_b$.}\vspace{2mm}
\begin{tabular}{|c||c|c|c|c||c|c|c|c|}
\hline - & \multicolumn{4}{|c||}{~~~LHC~($\sqrt S=14.$ TeV)~~~}&
\multicolumn{4}{|c|}{~~~TEVATRON~($\sqrt S=1.96$ TeV)~~~}\\
\hline\hline ~~~$Q^2$~~~ & ~~~$A$~~~ & ~~~$B$~~~ & ~~~$C$~~~ &
~~~$D$~~~
& ~~~$A$~~~ & ~~~$B$~~~ & ~~~$C$~~~ & ~~~$D$~~~ \\
\hline\hline $\sigma(^1P_1)(nb)$ & ~4.738~ & ~9.123~ & ~9.825~ &
~8.379~
& ~0.2555~ & ~0.6545~ & ~0.7547~ & ~0.5507~\\
\hline $\sigma(^3P_0)(nb)$ & 1.910 & 3.288 & 3.523 & 3.036
& 0.1161 & 0.2563 & 0.2966 & 0.2149\\
\hline $\sigma(^3P_1)(nb)$ & 4.117 & 7.382 & 7.304 & 6.682
& 0.2289 & 0.5597 & 0.6490 & 0.4780\\
\hline $\sigma(^3P_2)(nb)$ & 10.18 & 20.40 & 21.71 & 18.26
& 0.5096 & 1.350  & 1.515  & 1.102\\
\hline\hline
\end{tabular}
\label{tabq2} \vspace{-8mm}
\end{center}
\end{table}

From Tab.\ref{tabq2} about the total cross sections for the
hadronic production of $B_{cJ,L=1}^*$ at LHC and TEVATRON with the
different types of the choices, one may observe that the $Q^2$
dependence in the $P$-wave $B_{cJ,L=1}^*$ production is much more
stronger than that in the $S$-wave production. With the different
choices, the total cross sections of the production can be varied
so big as a factor $2.0\sim 3.0$.

For definiteness in the rest estimates on the production we will
adopt just Type $B$ for $Q^2$ choice.

\begin{figure}
\centering
\includegraphics[width=0.450\textwidth]{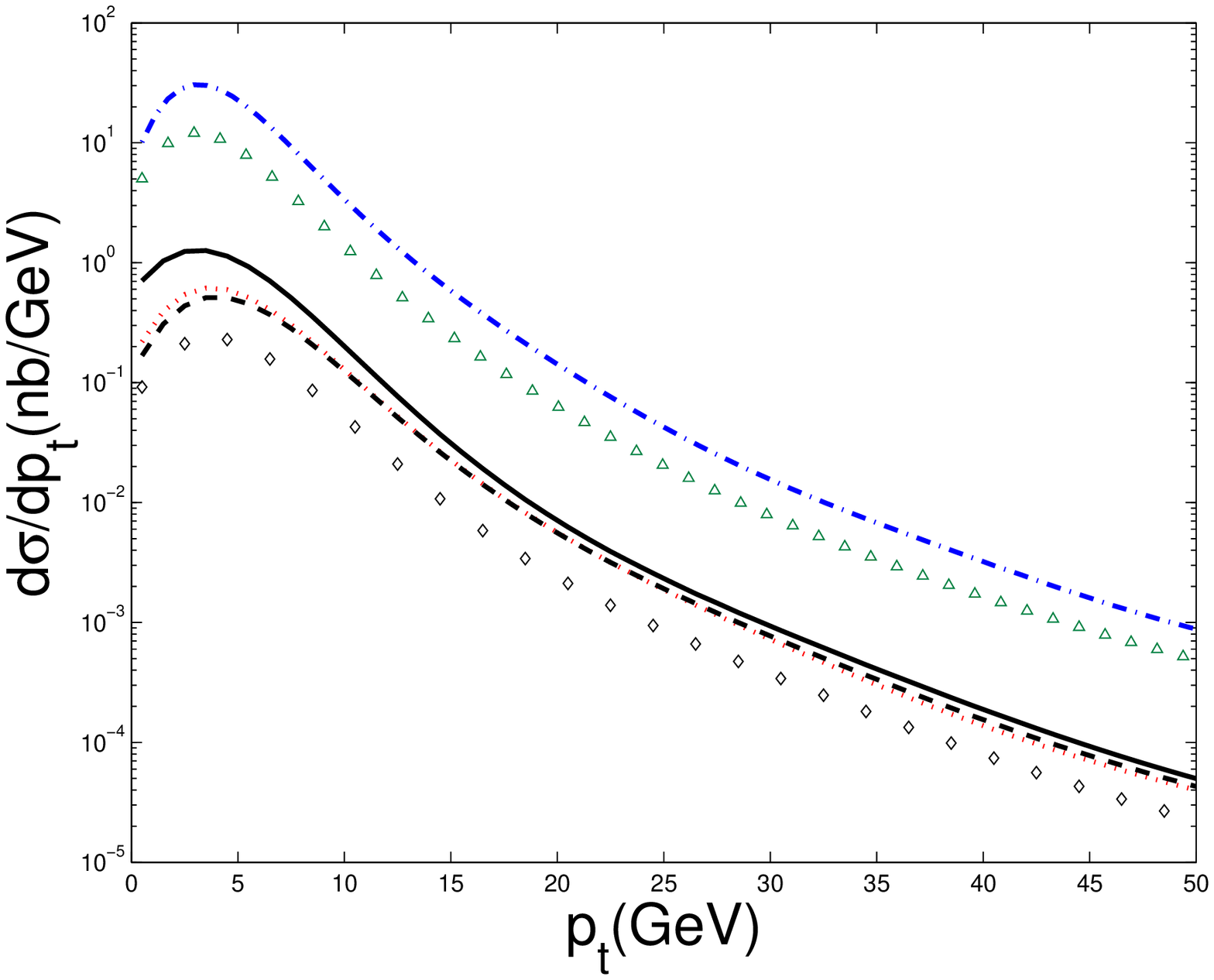}%
\includegraphics[width=0.450\textwidth]{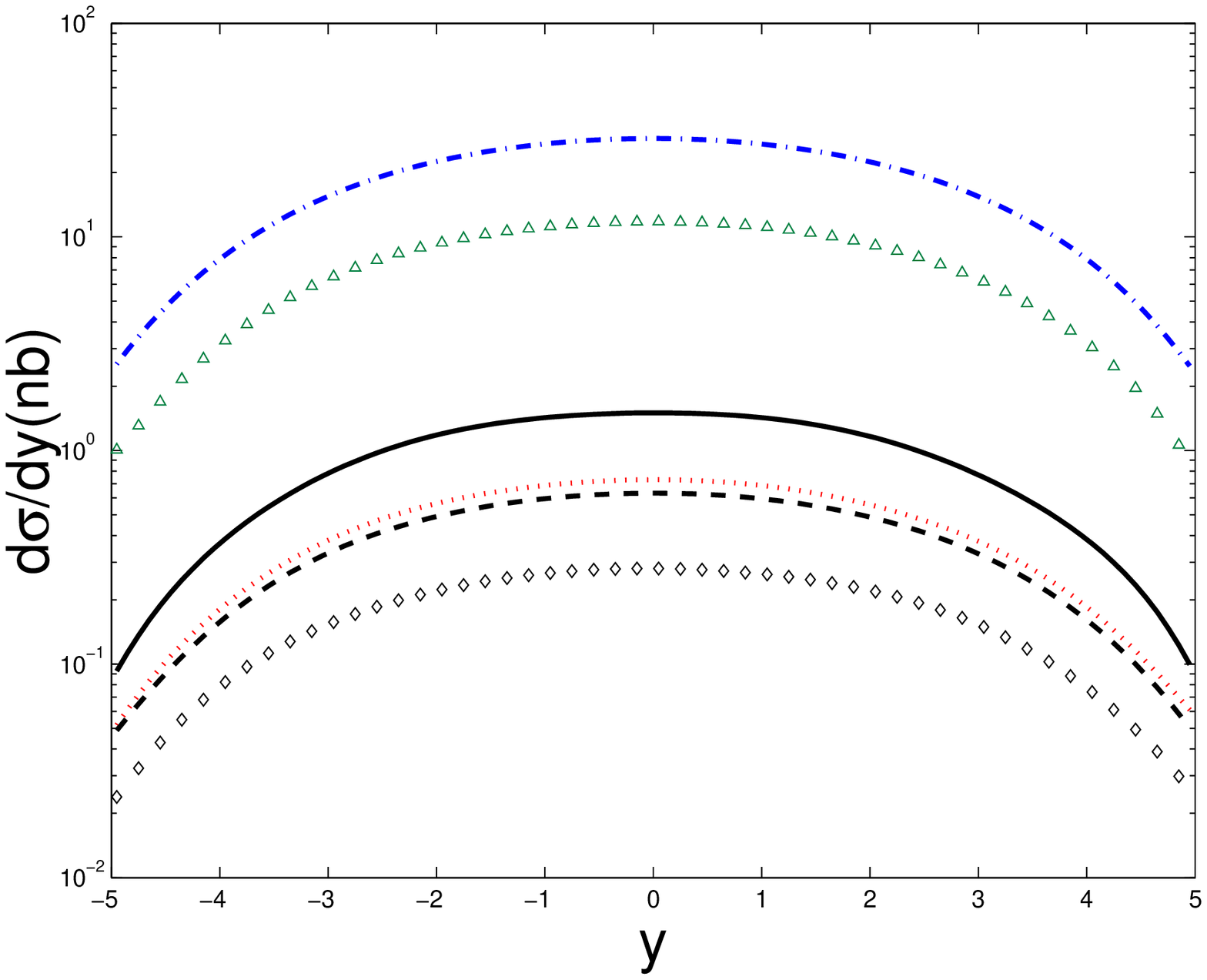}\hspace*{\fill}
\caption{\small The differential distributions of $p_t$ and $y$
for the hadronic production of $B_{cJ,L=1}^*$ at LHC. The dotted
line (down-middle), the diamond line (bottom), the dashed line
(next to bottom), the solid line (up-middle) are for $^1P_1$,
$^3P_0$, $^3P_1$, $^3P_2$, respectively. For comparison, the
results for $^1S_0$ and $^3S_1$ wave states are shown in triangle
line (next to top) and dash-dot line (top) correspondingly (with
parameters: $m_b=4.9$ GeV, $m_c=1.5$ GeV and
$m_{B_c}=m_{B_{cJ,L=1}^*}=m_c+m_b$).} \label{lhcwavep}
\vspace{-0mm}
\end{figure}

\begin{figure}
\centering
\includegraphics[width=0.44\textwidth]{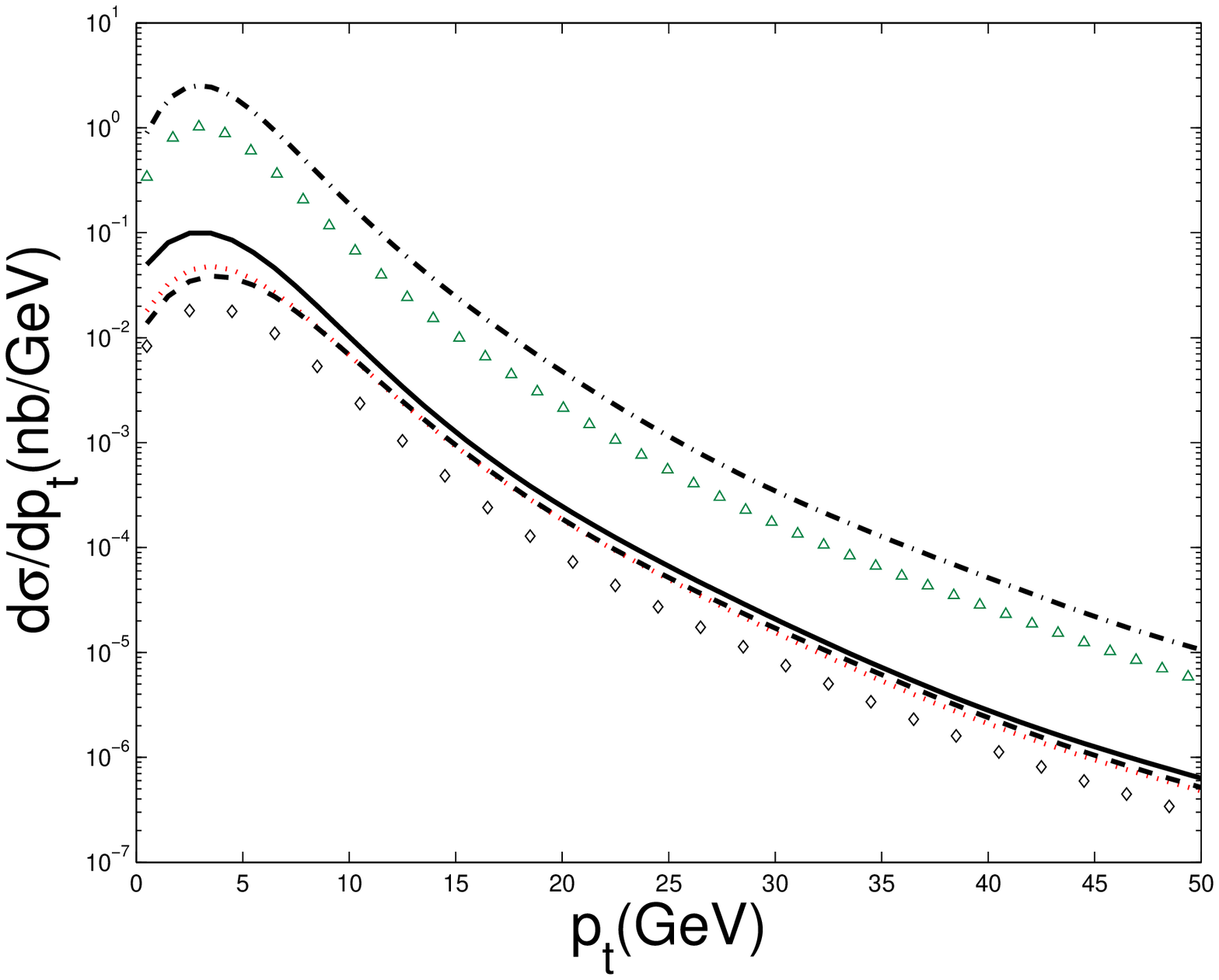}%
\includegraphics[width=0.450\textwidth]{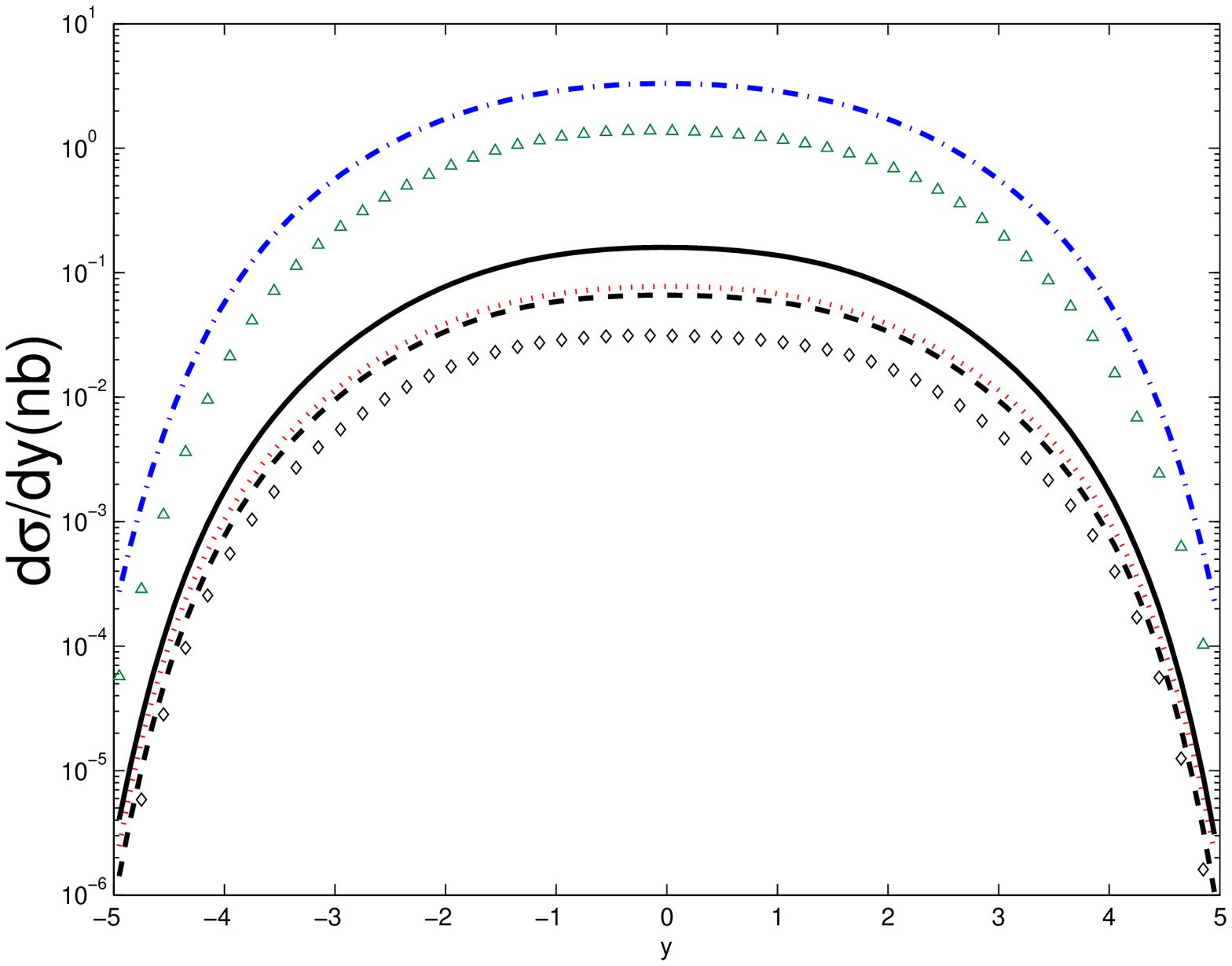}\hspace*{\fill}
\caption{The differential distributions of $p_t$ and $y$ for the
hadronic production of $B_{cJ,L=1}^*$ at TEVATRON. The dotted line
(down-middle), the diamond line (bottom), the dashed line (next to
bottom), the solid line (up-middle) are for $^1P_1$, $^3P_0$,
$^3P_1$, $^3P_2$, respectively. For comparison, the results for
$^1S_0$ and $^3S_1$ wave states are shown in triangle line (next
to top) and dash-dot line (top) correspondingly (with parameters:
$m_b=4.9$ GeV, $m_c=1.5$ GeV and
$m_{B_c}=m_{B_{cJ,L=1}^*}=m_c+m_b$).} \label{tevwavep}
\vspace{-0mm}
\end{figure}

For experimental references, we also calculate the distributions
of $p_t$ and $y$ (the $B_{cJ,L=1}^*$ transverse momentum and
rapidity) precisely and draw the curves on the distributions in
Fig.\ref{lhcwavep} for LHC and in Fig.\ref{tevwavep} for TEVATRON.

For comparison, in Figs.\ref{lhcwavep},\ref{tevwavep}, we also
show the $p_t$ and $y$ differential distributions for the
production of the S-wave states $^1S_0$ and $^3S_1$. One may
observe that the $p_t$ and $y$ dependence of the differential
distributions behaves quite similar to those of the $^1S_0$ state
production. The summed up $P$-wave production cross section of the
$P$-waves $^1P_1, ^3P_0, ^3P_1, ^3P_2$ is smaller than the
$S$-wave production, but it can reach to a fraction about $50\%$
of the $^1S_0$ production cross section.

\begin{figure}
\centering
\includegraphics[width=0.45\textwidth]{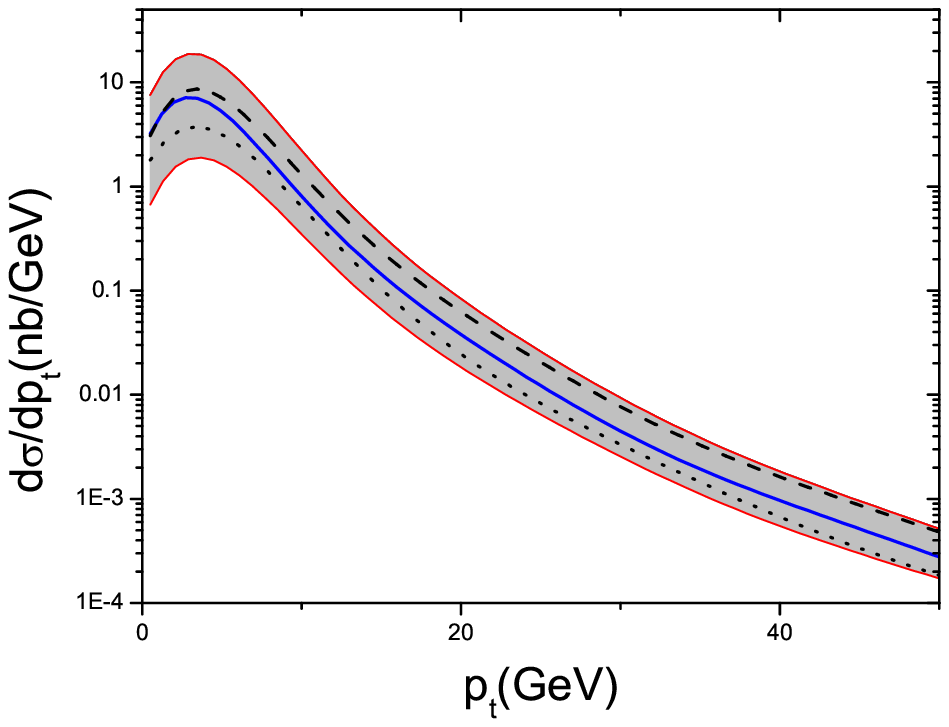}%
\includegraphics[width=0.45\textwidth]{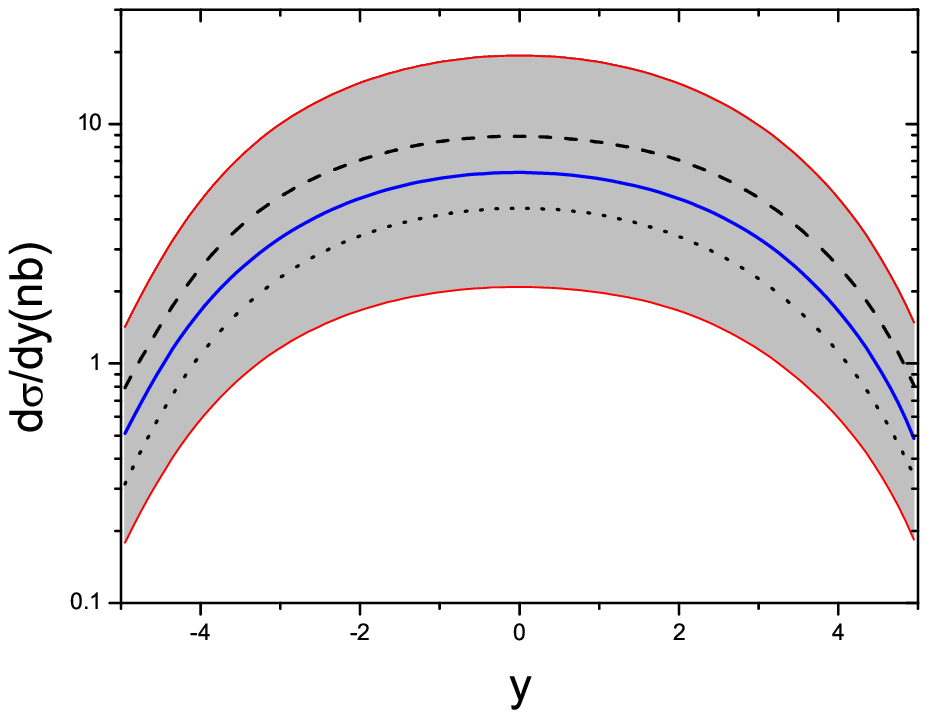}\hspace*{\fill}
\caption{The summed $p_t$- and $y$-distributions of all the
$P$-wave $B_{c,L=1}$ for different values of the factorization
scale $\mu^2_F$ and the renormalization scale $Q^2$ at LHC. The
upper edge of the band corresponds to
$\mu^2_F=4M_{t}^2,\;Q^2=M_{t}^2/4$ and the lower edge corresponds
to that of $\mu^2_F=M_{t}^2/4,\;Q^2=4M_{t}^2$. The solid line, the
dotted line and the dashed line corresponds to that of
$\mu^2_F=Q^2=M_{t}^2,\;$ $\mu^2_F=Q^2=4M_{t}^2,\;$
$\mu^2_F=Q^2=M_{t}^2/4$ respectively.} \label{diffQ2}
\vspace{-4mm}
\end{figure}
\begin{figure}
\centering
\includegraphics[width=0.45\textwidth]{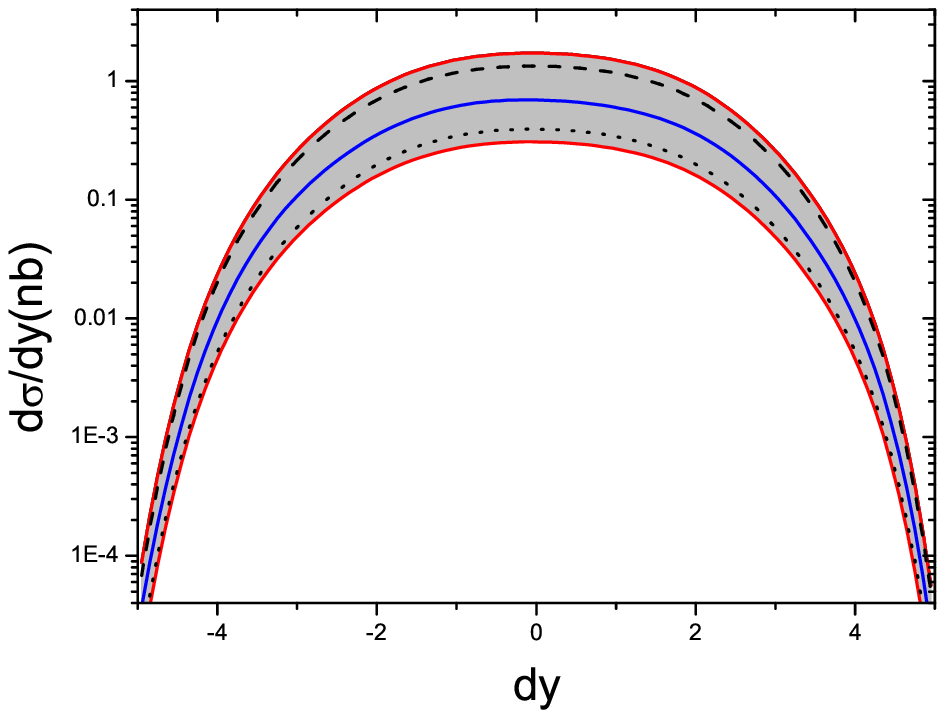}%
\includegraphics[width=0.45\textwidth]{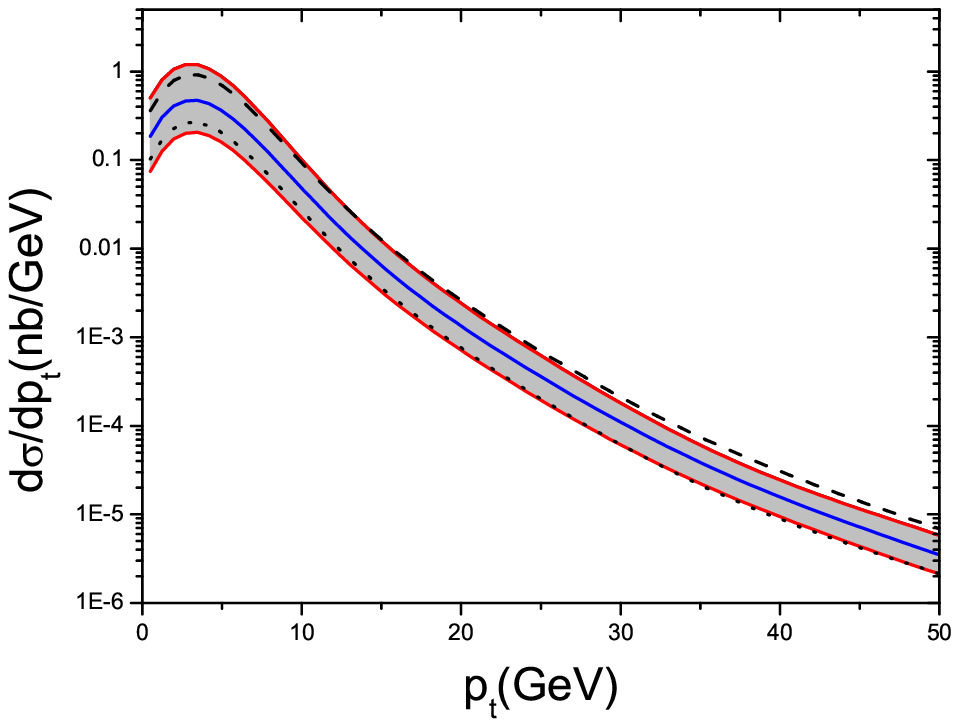}\hspace*{\fill}
\centering \caption{The summed up $p_t$- and $y$-distributions of
all the $P$-wave $B_{c,L=1}$ for different values of the
factorization scale $\mu^2_F$ and the renormalization scale $Q^2$
at TEVATRON.  The upper edge of the band corresponds to
$\mu^2_F=4M_{t}^2,\;Q^2=M_{t}^2/4$ and the lower edge corresponds
to that of $\mu^2_F=M_{t}^2/4,\;Q^2=4M_{t}^2$. The solid line, the
dotted line and the dashed line corresponds to that of
$\mu^2_F=Q^2=M_{t}^2$, $\mu^2_F=Q^2=4M_{t}^2$,
$\mu^2_F=Q^2=M_{t}^2/4$ respectively.}
\label{diffQ2tev}\vspace{-4mm}
\end{figure}

\begin{figure}
\centering
\hfill\includegraphics[width=0.44\textwidth]{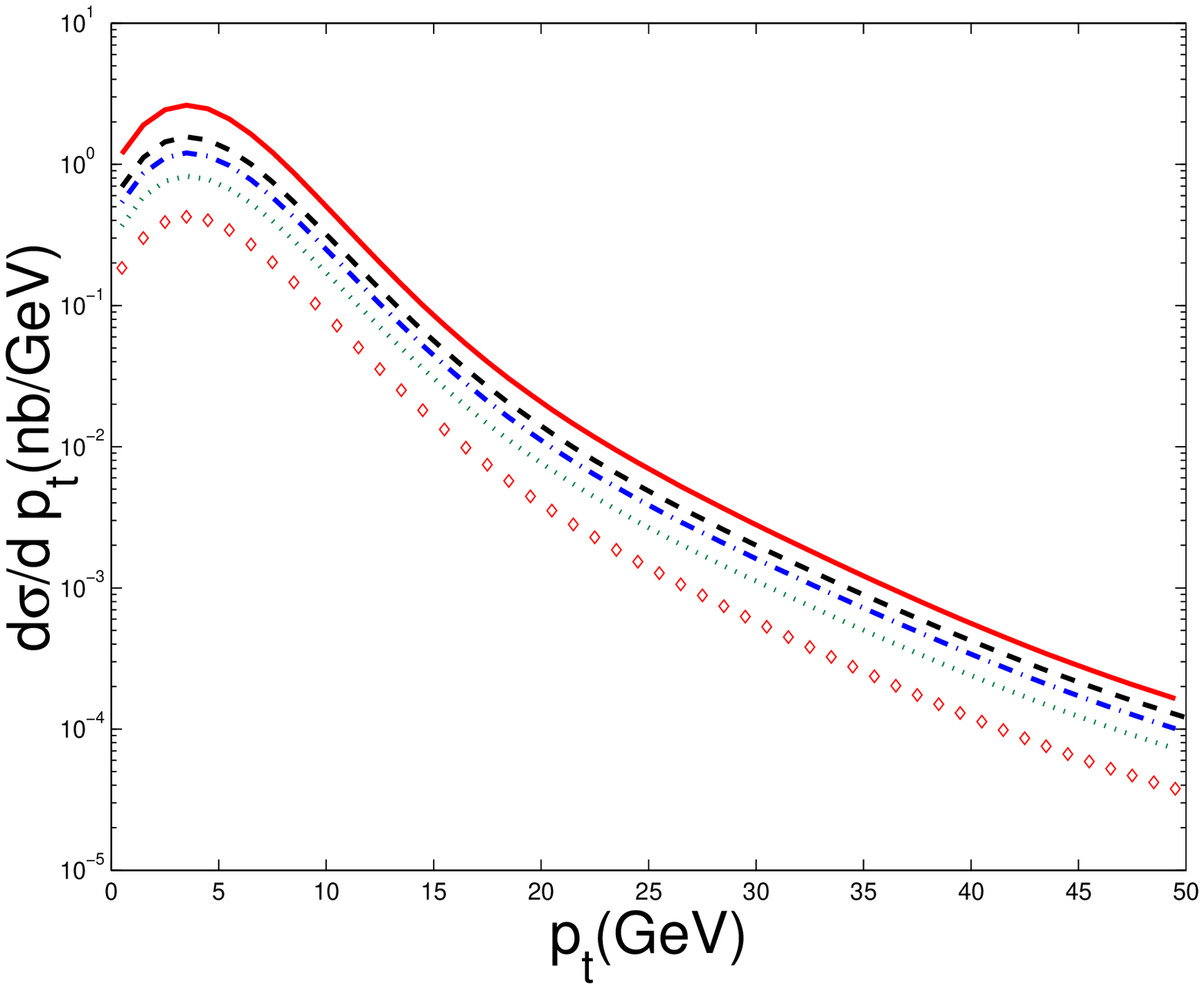}%
\includegraphics[width=0.45\textwidth]{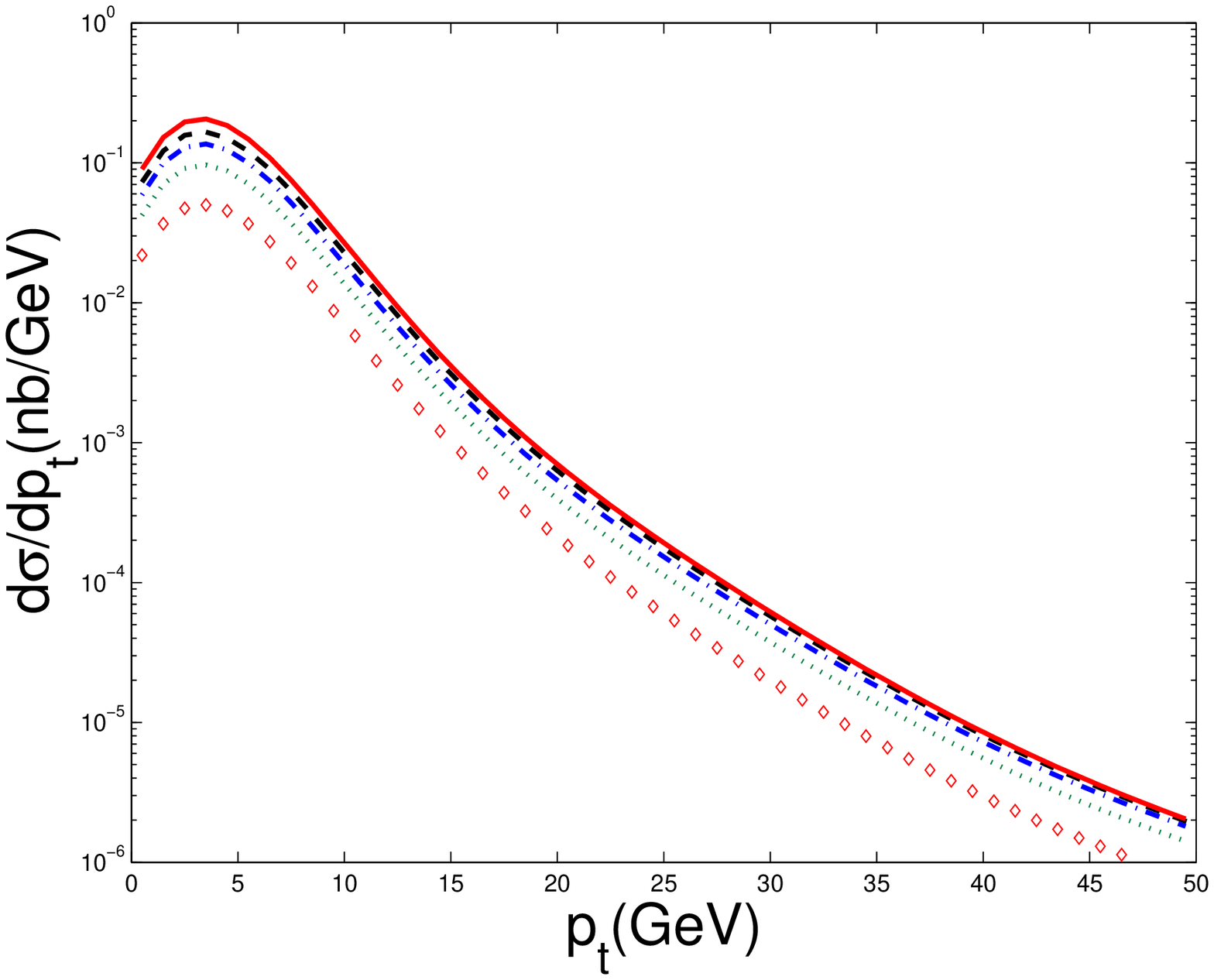}\hspace*{\fill}
\caption{The summed up differential distributions for all the
considered $P$-wave state production versus $p_{t}$ of
$B_{cJ,L=1}^*$ with a $y$ cut ($y_{cut}$) at LHC (left) and at
TEVATRON (right). Dashed line (next to top) with $y_{cut}=2.0$;
dash-dot line (middle) with $y_{cut}=1.5$; dotted line (next
bottom) with $y_{cut}=1.0$; diamond line (bottom) with
$y_{cut}=0.5$ and solid line (top) without $y_{cut}$.}
\label{lypt}
\end{figure}

\begin{figure}
\centering
\hfill\includegraphics[width=0.44\textwidth]{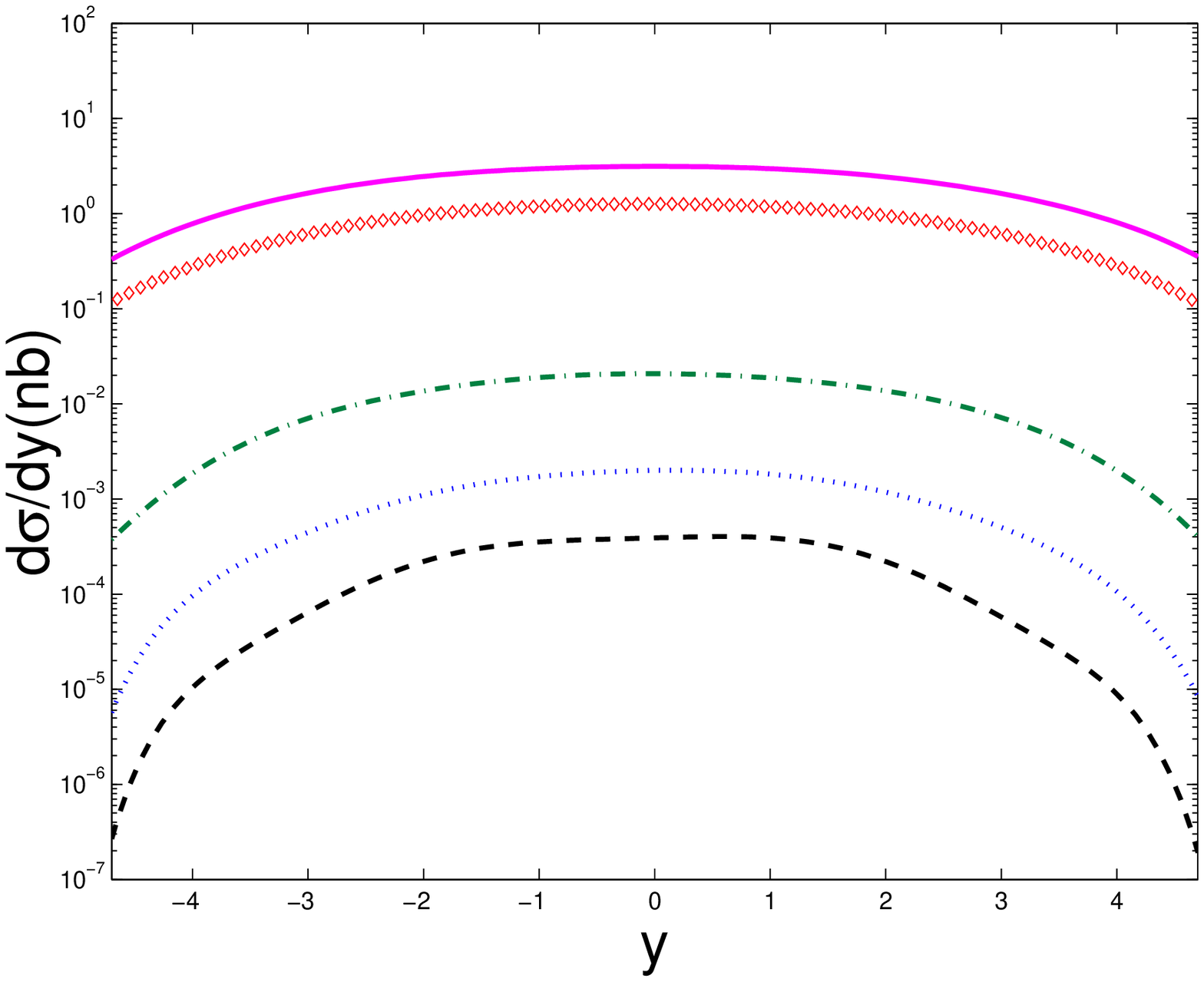}%
\includegraphics[width=0.45\textwidth]{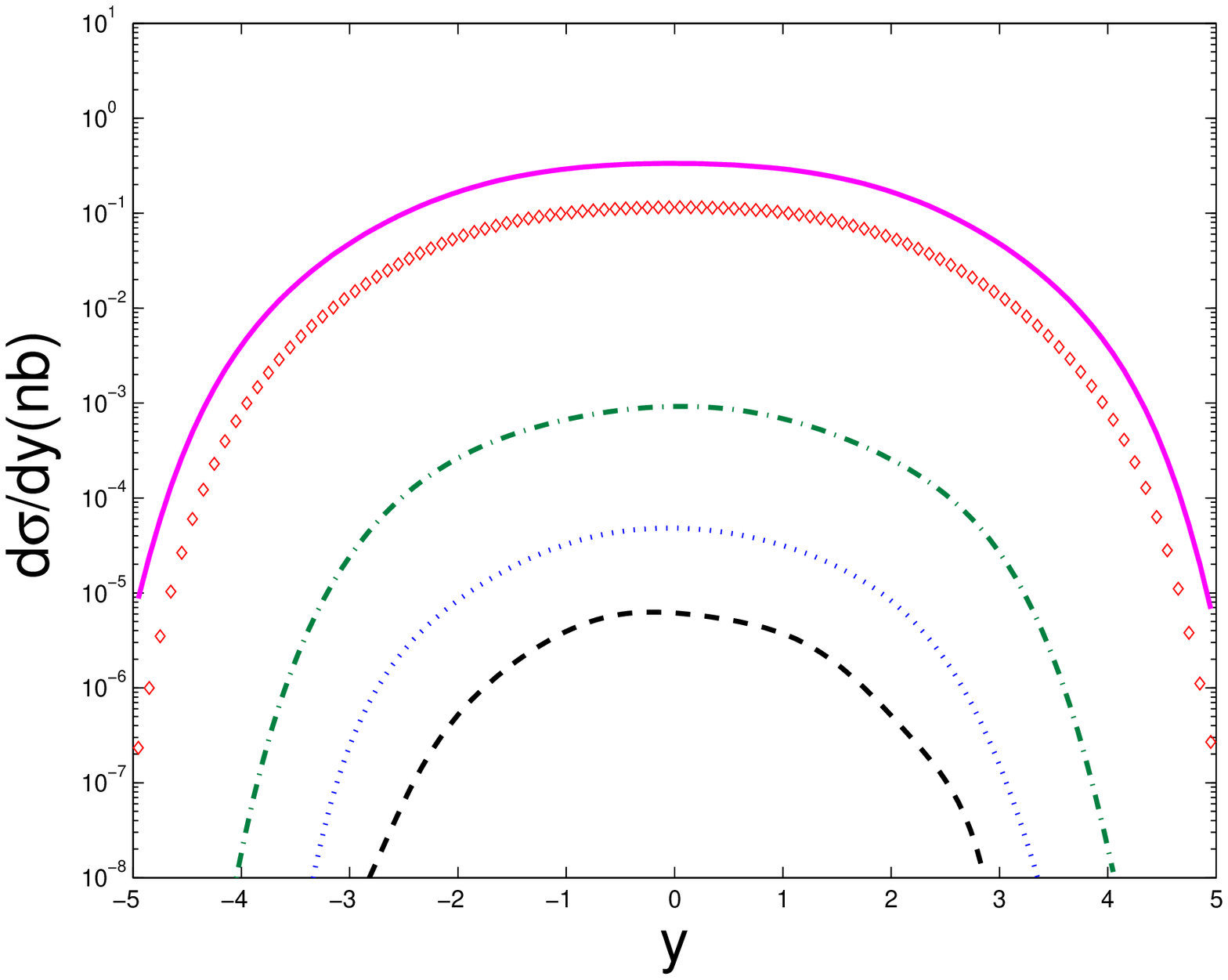}\hspace*{\fill}
\caption{\small The summed up differential distributions for all
the considered $P$-wave state production versus rapidity $y$ of
$B_{cJ,L=1}^*$ with a transverse momentum cut ($p_{tcut}$) at LHC
(left) and at TEVATRON (right). Diamond line (next to top) with
$p_{tcut}=5$ GeV; dash-dot line (middle) with $p_{tcut}=20$ GeV;
dotted line (next to bottom) with $p_{tcut}=35$ GeV; dashed line
(bottom) with $p_{tcut}=50$ GeV and solid line (top) with
$p_{tcut}=0$.} \label{lpty} \vspace{-4mm}
\end{figure}

\begin{figure}
\centering
\includegraphics[width=0.49\textwidth]{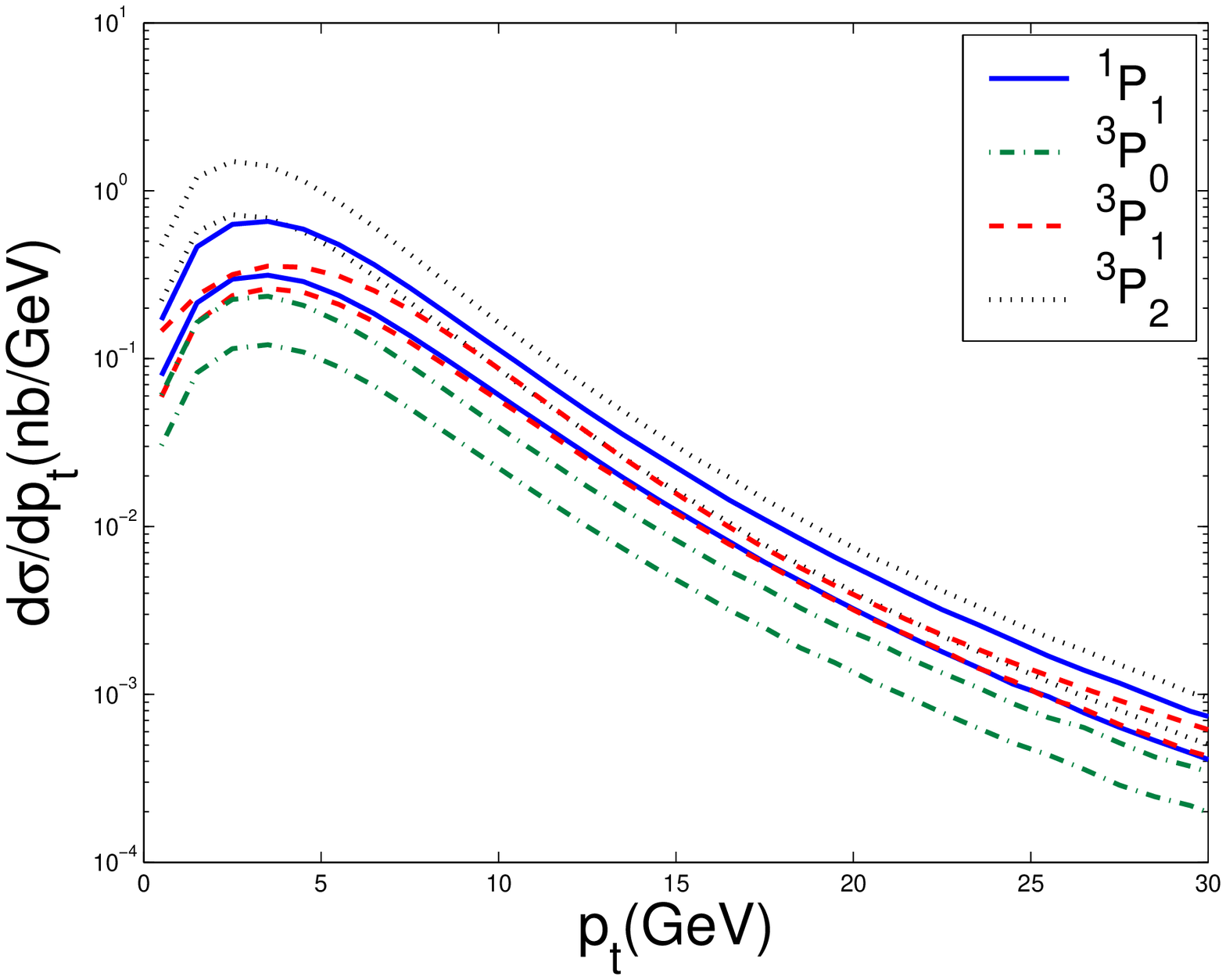}%
\includegraphics[width=0.49\textwidth]{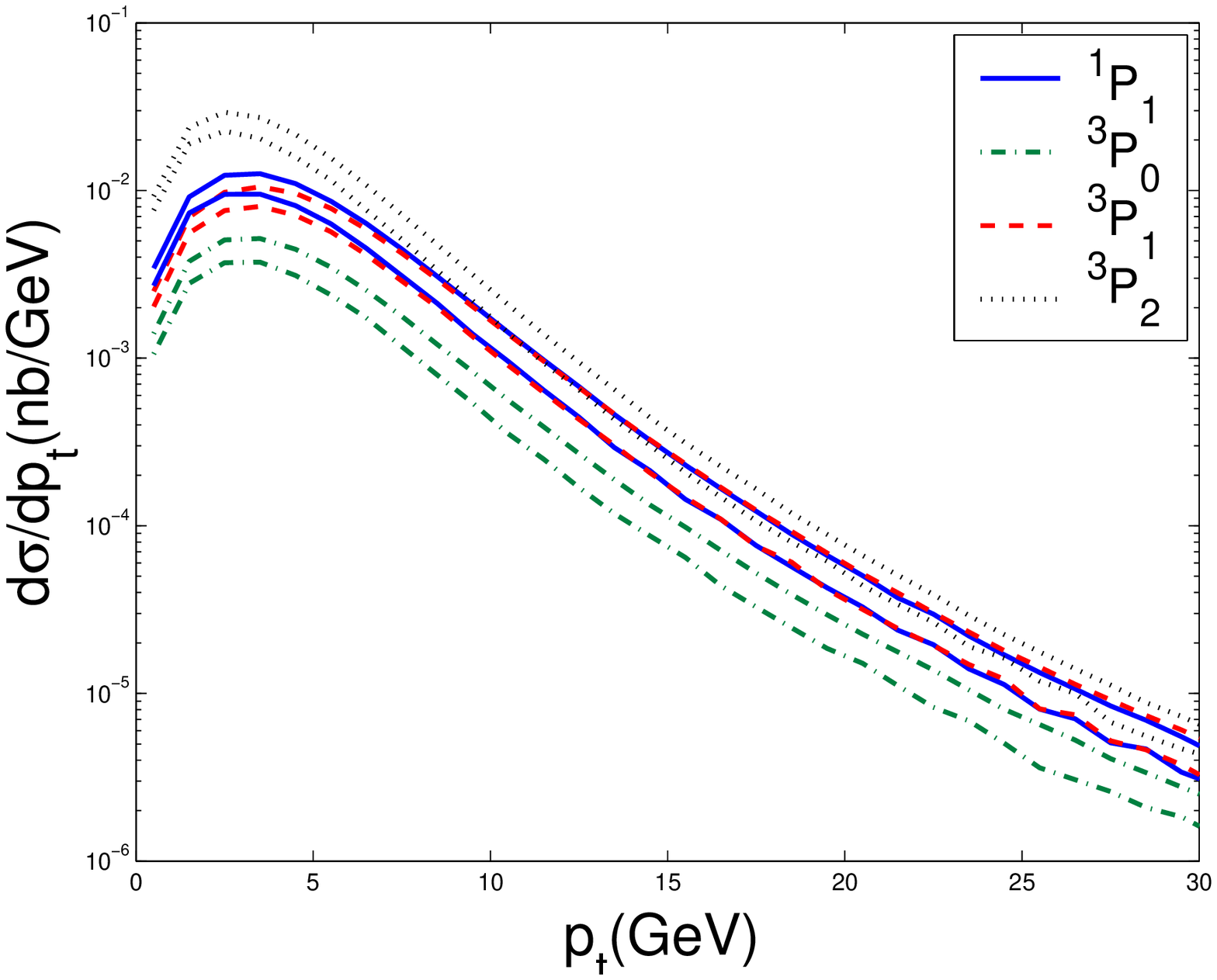}\hspace*{\fill}
\caption{\small The $p_t$ distributions of all the $P$-wave
$B^*_{c,L=1}$ ($^1P_1$ and $^3P_J\;(J=1,2,3)$) production at LHC
(left) and TEVATRON (right). Of the same type of lines, the upper
is for the choice: $m_b=4.9$ GeV, $m_c=1.5$ GeV, while the lower
is for the choice: $m_b=5.0$ GeV, $m_c=1.7$ GeV, respectively. For
LHC, we take $|y|\leq 1.5$ and for TEVATRON, we take $|y|\leq
0.6$.} \label{feynman0}
\end{figure}

From FIGs.\ref{lhcwavep},\ref{tevwavep}, one may also see that the
$p_t$ and $y$ differential distributions for the hadronic
production of $B_{cJ,L=1}^*$ are quite different at LHC and at
TEVATRON  due to the different C.M. energies of the two colliders.

To see the uncertainties from $Q^2$ choice, instead of variation
on the choices with $Q^2=\mu_F^2$, the authors in literature, such
as Ref.\cite{psi}, also try $Q^2\neq \mu_F^2$ in Eq.(\ref{pqcdf})
and see the uncertainty. Here following them, we calculate the
summed $p_t$- and $y$-distributions of all the $P$-wave
$B_{c,L=1}$ production for LHC and TEVATRON and present the
uncertainty on the distributions in
FIGs.\ref{diffQ2},\ref{diffQ2tev} respectively. One may see the
uncertainty can be so great as a factor $4.0$.

In high energy hadronic colliders, the events with a small $p_t$
and/or a large rapidity $y$ cannot be detected by detectors
directly, so for experimental studies and for precise purposes in
the estimates, only the events with proper kinematic cuts on $p_t$
and $y$ are taken into account. In Ref.\cite{changwu}, we did the
studies on the $S$-wave production, thus for experimental
references here we also try various cuts for the $P$-wave
production.

To study the cut effects on the production, here we consider the
production in summing up all of the $P$-wave ($^1P_0, ^3P_0,
^3P_1, ^3P_2$) production. Considering the abilities in measuring
rapidity of $B_c$ for the detectors CDF, D0 and BTeV at TEVATRON,
and ATLAS, CMS and LHC-B at LHC, we try the rapidity cuts
$y_{cut}\sim 1.5$ and higher for $p_t$ distribution and the
$p_{tcut}\sim 5$ GeV and higher for $y$ distribution. The results
with the four cuts: $y_{cut}=(0.5, 1.0, 1.5, 2.0)$ are put in
FIG.\ref{lypt} and the results with the four cuts: $p_{tcut}=5,
20, 35, 50$ GeV are put in FIG.\ref{lpty}.

From Fig.\ref{lypt}, we can see that the dependence of the $p_t$
distributions on $y_{tcut}$ for LHC is stronger than that for
TEVATRON. This is because that at TEVATRON, the sizable rapidity
distributions cover a smaller region in $y$, so generally speaking
the rapidity cuts affect the production stronger than that at LHC.
The correlations between $p_t$ and $y$ can be seen more clearly
from Fig.\ref{lpty}, the $y$-distribution various $p_t$-cuts. One
may observe that the dependence of the differential distribution
on rapidity $y$ with different $p_{tcut}$ is to exhibit a broader
profile at LHC than at TEVATRON.

We also would like specially to show the consequences due to the
approximation Eq.(\ref{bound}) with one by the two possibilities
described above, i.e. the one taken by ourselves:
$m_{B_{cJ,L=1}^*}=m_{B_c}=6.4$ GeV, $m_c=1.5$ GeV and $m_b=4.9$
GeV, and the other one taken by Refs.\cite{berezhnoy,berezhnoy2}:
$m_{B_{cJ,L=1}^*}=6.7$ GeV, $m_c=1.7$ GeV and $m_b=5.0$ GeV
quantitatively, because it is `fresh' in $P$-wave production. We
calculate $p_t$ distributions for the $P$-wave ($^1P_1, ^3P_0,
^3P_1, ^3P_2$) production with the two possibilities respectively
and draw the curves of the differential cross sections versus
$p_t$ in FIG.\ref{feynman0}. It is obvious that with our present
parameters for the constitute quarks, the distributions (hence the
total cross-sections) are about two times bigger than that of the
second choice for the constitute quark masses. Namely, the results
mean that the consequences due to the approximation
Eq.(\ref{bound}) are sizable and indicate improvement on the
approximation is requested.

\section{Discussions and summary}

In the paper, we have studied the hadronic production of the
$P$-wave $B_{cJ,L=1}^*$ in a way as analytical as possible, i.e.,
the amplitude of the hard subprocess $g+g\rightarrow
B^*_{c,L=1}+b+\bar{c}$ is reduced with the help of the FDC program
\cite{fdc} and the technique to expand each term of the amplitude
by the independent `bases' for the `elementary fermion strings'
analytically then to sum up all of the 36 terms (corresponding to
the 36 Feynman diagrams). Thus having the amplitude in expansion
of the bases squared, we have made the numerical calculations very
efficiently. The correctness for the amplitude and the program is
tested by demonstrating its gauge invariance numerically up to the
computer abilities and comparing the results for the subprocess
with those in Ref.\cite{berezhnoy} by taking the same parameters.
Having done the tests, we will add the whole program into the
$B_c$ meson generator as the new version {\bf BCVEGPY2.0}
soon\cite{bcvegpy2.0}.

From the figures FIGs.\ref{subwavep},\ref{lhcwavep},\ref{tevwavep}
and the tables Tabs.\ref{tabsub},\ref{tabq2}, one may see that, of
the hadronic $P$-wave production, that of $^3P_2$ is the biggest
and that of $^3P_0$ is the smallest. From the table
Tab.\ref{tabq2} and the figures FIGs.\ref{diffQ2},\ref{diffQ2tev},
one may see that, at LO, the uncertainties from the factorization
and renormalization energy scale choice, i.e. on $Q^2, \mu_F^2$
choices for the $P$-wave production, are quite great, even greater
than that of $S$-wave production. Furthermore, from the figure
FIG.\ref{feynman0}, one may see that the approximation
Eq.(\ref{bound}) causes a great uncertainty in the production
estimate.

Relating to the approximation, our results with the choice
$m_{B_{cJ,L=1}^*}=m_{B_c}=6.4$ GeV, $m_c=1.5$ GeV and $m_b=4.9$
GeV can be almost two times bigger than the results obtained in
Ref.\cite{berezhnoy} with the choice $m_{B_{cJ,L=1}^*}=6.7$ GeV,
$m_c=1.7$ GeV and $m_b=5.0$ GeV at LHC and TEVATRON. The fact
about the factor $2$ can be understandable: as known, smaller
quark masses will cause the cross section of the four quark
production $gg\to c+\bar{c}+b+\bar{b}$ bigger and they will also
cause an increase effect in the binding $c+\bar{b}\to
B_{cJ,L=1}^*$, thus when we take smaller masses of quarks and also
the bound state mass $M=m_b+m_c$ in the meantime, we obtain such
bigger results by the factor $2$ than those in
Ref.\cite{berezhnoy}. We think that for the production the true
values (without the approximation Eq.(\ref{bound}) for a more
accurate estimate) should be smaller than those we present here,
due to a comparatively small value of the masses of $B_{cJ,L=1}^*$
taken here, but the true values should be bigger than those
obtained in Ref.\cite{berezhnoy} due to there comparatively big
values for the masses of $c$ and $b$ quarks taken for the factor
of the four quark production inside the $P$-wave production
respectively. Since the factor of the involved production $gg\to
b\bar{b}c\bar{c}$ is guaranteed in gauge invariance, here only the
factor $c\bar{b}\to B_{cJ,L=1}^*$ is not well treated in the
present results, while in Ref.\cite{berezhnoy} not only the factor
$c\bar{b}\to B_{cJ,L=1}^*$ but also the factor of the involved
four quark production are not well-treated by comparatively bigger
values for the quarks and the bound states, so we suspect that the
true values for the production might be closer to the presented
ones here, rather than those in Ref.\cite{berezhnoy}. Note that
the term $B^{\lambda'\;\mu}_{ss_z}q_\mu$ in Eq.(\ref{eq:ab}) does
not play any role for $S$-wave $B_c$ and $B_c^*$ production at
all, thus for the $S$-wave production the approximate choices
about the values of $P=q_{c1}+q_{b2}\,,\; m_c\,,\; m_b$ and
$m_{B_c(B_c^*)}$ are not so sensitive as the $P$-wave production.

The summed total cross section for all of the considered hadronic
$P$-wave production can be so big as a half of the direct
production of the ground state $B_c(^1S_0)$. Considering the the
fact that almost all of the $P$-excited states $B_c^*(^3S_1)$
decay to the ground state $B_c$, the contribution from the
$B_c^*(^3S_1)$ production to the `final $B_c$-wave production', we
may conclude that the contribution can be about $20\%$ in total.
From the $p_t$ and $y$ distributions, one may see that $P$-wave
and $S$-wave production behave quite similar, that no matter what
$p_{tcut}$ and $y_{cut}$ are chosen, the $P$-wave production
itself with such big cross sections is worth while to study the
possibility directly measuring the $P$-wave $B_c$-states
seriously. Especially, the properties of the $P$-wave mesons
$B^*_{(c,L=1)}$ are crucial in understanding the mass spectrum of
the $c\bar{b}$-quarkonium states and testing the potential models.

\vspace{20mm} \noindent {\Large\bf Acknowledgements:}
This work was supported in part by the Natural Science Foundation
of China (NSFC).\\

\appendix
\section{The linear polarization vector and tensor}

As stated in the text, the linear polarization vector (tensor)
contains real number only if the meson (bound state) is a vector
(tensor) meson, so the linear ones are better than complex
circular polarization ones in numerical calculations, thus the
linear explicit expressions for them are quite useful in writing
numerical programs. Hence, we present the precise expressions for
the linear polarization vector if the meson is a vector one, and
the polarization tensor if the meson is a tensor one in the
appendix.

We set a coordinate first, in which $z$-axial is in the direction
of the incoming hadron, and the momentum of the concerned meson
(bound state) in the direction:
$P^\mu=(P_0,\;|\vec{P}|\sin\theta\cos\phi,\;
|\vec{P}|\sin\theta\sin\phi,\;|\vec{P}|\cos\theta)$ with
$|\vec{P}|=\sqrt{P_0^2-M^2}$, i.e., $\theta$ is the polar angle
and $\phi$ is the azimuth angle in the coordinate. The space-like
linear polarization vector for the meson can be expressed as:
\begin{eqnarray}
\epsilon^\mu_{x}(\vec{P})&:&(0,\;\cos\theta\cos\phi,\;\cos\theta\sin\phi,\;-\sin\theta)\\
\epsilon^\mu_{y}(\vec{P})&:&(0,\;-\sin\phi,\;\cos\phi,\;0)\\
\epsilon^\mu_{z}(\vec{P})&:&\frac{1}{M}(|\vec{P}|,\;P_0\sin\theta\cos\phi,
\;P_0\sin\theta\sin\phi,\; P_0\cos\theta)\,.
\end{eqnarray}
As the request of the polarization vector definition, they do
satisfy the conditions:
\begin{equation}
\epsilon_{i}\cdot P=0\,,\,\;\;\; \epsilon_{i} \cdot
\epsilon_{j}=-\delta_{ij}\,,\,\,\;\;\;(i,j=x,y,z)\,.
\end{equation}

Accordingly, the `linear polarization tensor'
$\epsilon^{\alpha\beta}_{J_z}(P)$ for $B_c^*(^3P_2)$ can be
constructed in the following way i.e. the five components
($J_z=1,2,\cdots 5$) of the polarization tensor are constructed in
terms of $\epsilon_{x}(\vec{P})$, $\epsilon_{y}(\vec{P})$ and
$\epsilon_{z}(\vec{P})$:
\begin{eqnarray}
&\epsilon^{\mu\nu}_{1}=\frac{1}{\sqrt{2}}(\epsilon_x^{\mu}
\epsilon_y^{\nu}+\epsilon_y^{\mu} \epsilon_x^{\nu})\,,\;\;\;\;
\epsilon^{\mu\nu}_{2}=\frac{1}{\sqrt{2}}(\epsilon_x^{\mu}
\epsilon_z^{\nu}+\epsilon_z^{\mu} \epsilon_x^{\nu})\,,\nonumber \\
&\epsilon^{\mu\nu}_{3}=\frac{1}{\sqrt{2}}(\epsilon_y^{\mu}
\epsilon_z^{\nu}+\epsilon_z^{\mu} \epsilon_y^{\nu})\,,\;\;\;\;
\epsilon^{\mu\nu}_{4}=\frac{1}{\sqrt{2}}(\epsilon_x^{\mu}
\epsilon_x^{\nu}-\epsilon_y^{\mu} \epsilon_y^{\nu})\,,\nonumber\\
&\epsilon^{\mu\nu}_{5}=\frac{1}{\sqrt{6}}(\epsilon_x^{\mu}
\epsilon_x^{\nu}+\epsilon_y^{\mu} \epsilon_y^{\nu}-
2\epsilon_z^{\mu} \epsilon_z^{\nu})\;,
\end{eqnarray}
where again $\mu$ and $\nu$ are the Lorentz vector indexes. Again,
one may easily check that the polarization tensor components
satisfy the conditions:
\begin{eqnarray}
&&\epsilon^{\dagger}_{i}.\epsilon_{j}=-\delta_{ij}\;\;,
\epsilon_{i\mu}^{\mu}=g_{\mu\nu}\epsilon_{i}^{\nu\mu}=0\;\;,
P_\mu\cdot \epsilon^{\mu\nu}_{i}=\epsilon_{i\nu\mu}\cdot
P^\mu=0\;\;\; {\rm for}\;\; (i,j=1,\cdots,5).
\end{eqnarray}

\end{document}